\documentclass[%
 reprint,
superscriptaddress,
showpacs,preprintnumbers,
 amsmath,amssymb,
 aps,
]{revtex4-1}
\usepackage{graphicx}
\usepackage{dcolumn}
\usepackage{bm}%
\usepackage[colorlinks=true,citecolor=blue]{hyperref}
\DeclareUnicodeCharacter{2032}{\ensuremath{'}}
\hypersetup{colorlinks=true,citecolor=blue,linkcolor=red,urlcolor=blue}

\usepackage{float}
\usepackage{amsmath}

\usepackage{bm}
\newcommand{\beq}{\begin{equation}}
\newcommand{\eeq}{\end{equation}}
\newcommand{\beqnar}{\begin{eqnarray}}
\newcommand{\eeqnar}{\end{eqnarray}}
\newcommand{\bfig}{\begin{figure}}
\newcommand{\efig}{\end{figure}}

\usepackage{color}
\usepackage[dvipsnames]{xcolor} 
\usepackage{changes}

\begin{document}
\title{Dynamical nonlinear optical response in time-periodic quantum systems}

\author{S. Sajad Dabiri}
\affiliation{Department of Physics, Shahid Beheshti University, 1983969411 Tehran, Iran}
\author{Reza Asgari}
\affiliation{Department of Physics, Zhejiang Normal University, Jinhua 321004, China}
\affiliation{School of Physics, Institute for Research in Fundamental Sciences (IPM), Tehran 19395-5531, Iran}
\date{\today}
\vspace{1cm}
\newbox\absbox
\begin{abstract}
We present a comprehensive theoretical framework for calculating the linear and nonlinear optical responses of time-periodic quantum systems. Using density matrix evolution in the Floquet basis and adopting the length gauge, our approach incorporates both interband and intraband contributions of the position operator, enabling detailed insights into photon-assisted transitions and their associated optical phenomena. Notably, we identify a divergent ac response to dc fields in Floquet systems, reminiscent of a Drude peak at finite frequencies. This framework generalizes to optical tensor conductivity calculations at arbitrary perturbation orders and captures various DC photocurrents, including shift current, injection current, and Berry dipole contributions, under specific limits. To demonstrate the versatility of our method, we compute linear and nonlinear optical conductivities for one- and two-dimensional systems, revealing phenomena such as band inversions, j-photon-assisted transitions, and high harmonic generation. These results highlight the interplay between periodic driving and nonlinear optical effects, offering different avenues for exploring dynamical topological properties and their applications in ultrafast spintronics, optoelectronics, and strongly correlated systems. Our findings provide a robust platform for analyzing the complex optical behavior of driven quantum systems and guiding experimental investigations in this rapidly evolving field.
\end{abstract}
\maketitle

\section{Introduction}
{The nonlinear response enables us to uncover new physics not apparent in linear regimes and can be a very useful classification tool to study materials.} 
Recent advancements have significantly deepened our understanding of linear and nonlinear responses in real systems \cite{Boyd2007, shen2018linear}, particularly within the contexts of topological phases \cite{weinberg2017adiabatic, harper2020topology, PhysRevLett.129.040402, PhysRevLett.116.016802, PhysRevB.108.235406}, twisted bilayer graphene \cite{PhysRevResearch.1.023031, PhysRevLett.132.166901}, and open quantum systems \cite{mori2023floquet}. These discoveries have profound implications for quantum engineering, ultrafast spintronics, and the characterization of topological edge states \cite{mochizuki2020stability} in periodically driven quantum systems \cite{newoptic}. 
Periodically driven quantum systems, also known as Floquet systems, provide a rich platform for exploring non-equilibrium phenomena \cite{berdanier2018floquet}. These systems exhibit unique properties such as non-thermal behavior\cite{PhysRevB.97.014311} and host a variety of topologically non-trivial phases \cite{berdanier2018floquet}. Additionally, they enable the realization of exotic states, including Floquet time crystals. {The interplay between periodic driving and nonlinear effects unlocks phenomena such as high harmonic generation, photocurrents, and stability transitions of topological edge states, paving the way for innovative applications in optoelectronics and strongly correlated systems \cite{oka2019floquet, rudnerreview}.}

 Nonlinear transport phenomena, characterized by deviations from linear response under intense external fields, have garnered particular attention due to their ability to reveal novel effects. 
 {For instance, injection currents \cite{day2024nonperturbative} and second harmonic generation \cite{shan2021giant} in Floquet systems have been experimentally observed, 
demonstrating the interplay of band topology and nonlinear optical processes. Despite this progress, }understanding the dominant nonlinear responses in Floquet systems and their dependence on optical processes, band structure, and interband dynamics remains an open challenge
  {and can be harnessed to create novel optical devices}. The nonlinear response in time-independent real materials \cite{chu2004beyond, di2024all} remains an active research area with important implications for controlling quantum dynamics. { The Floquet theorem can be used to understand the topological nonlinear effects in solids \cite{morimoto2016topological,matsyshyn2021rabi}. Recently, the Floquet formulation of the nonlinear response has been developed for static systems \cite{popova2024microscopic} especially for applications of functional density theory \cite{alliati2023floquet}. }

There are two primary formulations in the literature for calculating nonlinear conductivity from the Hamiltonian of static systems: the velocity gauge and the length gauge, each with its own advantages and disadvantages \cite{taghi2017, di2019resolution}. {Both the velocity gauge and the length gauge are comparable methods for determining a system's optical response and have been thoroughly documented in the literature. Depending on the system and context, the gauge dependence arguments in the Floquet systems provide distinct viewpoints and computing efficiencies, but are mathematically similar. The electric field and the charged particles' spatial location are directly coupled by the length gauge.  The velocity gauge, however, relates the vector potential to the particle's momentum, making it especially comprehensible for systems where the dipole moment or polarization is important. It is useful for systems analysis where angular momentum or magnetic field interactions are important. As a result, while the velocity gauge offers a momentum-space perspective, the length gauge offers direct intuition on the effect of the electric field on dipoles. Furthermore, certain symmetries or boundary conditions may be simpler to manage in one gauge than the other. Therefore, the physical properties of the issue, selected computational framework, and numerical stability considerations all influence the decision between the length and velocity gauges.}

Recently, the velocity gauge approach has been successfully applied using Feynman diagrams \cite{parker2019}. However, this method often suffers from spurious divergences at low frequencies, which can be problematic in numerical calculations with a limited number of bands \cite{taghi2017}. In contrast, the length gauge, which reveals intriguing effects such as gyration current and Fermi surface phenomena \cite{watan2021}, faces challenges due to the complexity of the matrix elements of the position operator. {Moreover, in order to find the matrix elements of the position operator a smooth gauge for the phase of Bloch wave functions must be chosen across the Brillouin zone.}

The linear response of Floquet systems has been extensively studied in previous works \cite{oka, wu2011, deh2015, seradjeh2020,tiwari2023floquet,gu2018optical}. Notably, the time dependence of linear conductivity has been addressed within the velocity gauge formulation \cite{wu2011, seradjeh2020}. Despite significant advancements, key questions remain unanswered regarding the nonlinear responses of Floquet systems, for instance: \textit{What are the dominant nonlinear responses to time-dependent quantum systems? How is it influenced by the bands and different optical processes, and what information can it provide about interband processes?} To address these questions, this work develops a comprehensive framework to calculate linear and nonlinear optical conductivities of time-periodic quantum systems. Utilizing density matrix evolution in the Floquet basis and adopting the length gauge, we construct a versatile formalism capable of describing optical tensor conductivity at arbitrary perturbation orders. This approach captures phenomena such as divergent AC responses to DC fields, reminiscent of Drude peaks, as well as various DC photocurrents, including shift currents, injection currents, and Berry dipole contributions.

{We apply this framework to one- and two-dimensional systems, unveiling features such as j-photon-assisted transitions, band inversions, and dynamical gaps in optical conductivities. These results highlight the intricate connections between periodic driving, nonlinear responses, and band topology, offering valuable insights into the design of advanced quantum materials and devices. Our work provides a robust platform for exploring the rich physics of driven quantum systems, with broad applications in ultrafast optics, quantum information science, and condensed matter physics.}

\section{Perturbation theory}
We consider a system with dynamics defined by Hamiltonian $H({\bf k},t)$. For time periodic Bloch Hamiltonian $H(t+T)=H(t)$ (where momentum index ${\bf k}$ is suppressed) with time period $T=2\pi /\Omega$ the Floquet theorem suggests the form of Schr\"{o}dinger equation eigenvectors as (assuming $e=\hbar =1$ hereafter): \cite{floquetref}  
$|{{\psi }_{\alpha }}(t)\rangle ={{e}^{-i{{\epsilon }_{\alpha }}t}}|{{\phi }_{\alpha }}(t)\rangle
$
where Greek letter $\alpha$ is the band index, $|{{\phi }_{\alpha }}(t)\rangle =|{{\phi }_{\alpha }}(t+T)\rangle$ are time periodic Floquet quasi modes and ${{\epsilon }_{\alpha }}$ is quasienergy which is defined modulo $\Omega$, i.e. ${{\epsilon }_{\alpha }}\cong {{\epsilon }_{\alpha }}+\Omega $ and the momentum dependence is implicit. We restrict ${{\epsilon }_{\alpha }}$ to be in the first Floquet Brillouin zone i.e. $-\frac{\Omega}{2}<{{\epsilon }_{\alpha }}\le \frac{\Omega}{2}$.

Later, we construct a perturbation theory to predict the density matrix after applying a potential $\lambda V(t)$ to the time-dependent system. The time evolution of the density matrix is derived from the von Neumann equation:
\begin{equation}
i{{\partial }_{t}}\rho ({\bf k},t)=[H({\bf k},t)+\lambda V(t),\rho ({\bf k},t)],
\end{equation}
where symbol $[A,B]=AB-BA$ is the commutator.  Assuming $\lambda$ as a small constant, we make an expansion for $\rho $ in powers of $\lambda$ as $\rho ={{\rho }^{[0]}}+\lambda {{\rho }^{[1]}}+{{\lambda }^{2}}{{\rho }^{[2]}}+...$.  We assume that the density matrix is diagonal in the Floquet basis i.e. $
{{\rho }^{[0]}}=\sum\limits_{\eta }{{{f}_{\eta }}|{{\phi }_{\eta }}(t)\rangle \langle {{\phi }_{\eta }}(t)|}
$
with occupation of Floquet states ${{f}_{\eta }}$ remaining constant in time.   {This is justified if the drive was turned on in the distant past \cite{privitera2017non}, allowing all observables to synchronize with the drive and the system to enter the Floquet diagonal ensemble.} As shown in \cite{Note2}, by defining $\rho _{\alpha \beta }\equiv \langle {{\phi }_{\alpha }}(t)|{{\rho }}|{{\phi }_{\beta }}(t)\rangle $, and $\epsilon_{\alpha \beta }=\epsilon_{\alpha}-\epsilon_{\beta } $, there is a recursion relation between density matrix elements of different orders as:
\begin{equation}
\begin{aligned}
\rho _{\alpha \beta }^{[n]}=-i{{e}^{-i{{\epsilon }_{\alpha \beta }}t}}\int_{-\infty }^{t}{{{e}^{i{{\epsilon }_{\alpha \beta }}t'}}\langle {{\phi }_{\alpha }}({{t}^{\prime }})|[V({{t}^{\prime }}),{{\rho }^{[n-1]}}]|{{\phi }_{\beta }}({{t}^{\prime }})\rangle d{{t}^{\prime }}}
\end{aligned}
\label{rho1int}
\end{equation}
where $n\in \mathbb{N}$. In the following, we find the optical tensor conductivity in the length gauge i.e. $V(t)=\mathbf{r}\cdot\mathbf{E}e^{-i\omega_1 t}$ where $\mathbf{r}$ and $\mathbf{E}$ are the position operator and amplitude of the electric field.
It is important to note that the matrix elements of $[{\bf r},\rho]$ when evaluated based on Bloch states, require careful consideration. The ${\bf r}$ can be divided into interband and intraband contributions $\mathbf{r}=\mathbf{r}^i+\mathbf{r}^e$ \cite{sipe1995} and the matrix elements of the position operator between Bloch states generalized to Floquet system are:   
\begin{equation}
\begin{aligned}
  & \langle {{\phi }_{\alpha }}(\mathbf{k},t)|{{\mathbf{r}}^{i}}|{{\phi }_{\beta }}({{\mathbf{k}}^{\prime }},t)\rangle ={{\delta }_{\alpha \beta }}[\delta (\mathbf{k}-{{\mathbf{k}}^{\prime }}){{\xi }_{\alpha \alpha }}+i{{\partial }_{\mathbf{k}}}\delta (\mathbf{k}-{{\mathbf{k}}^{\prime }})] \\ 
 & \langle {{\phi }_{\alpha }}(\mathbf{k},t)|{{\mathbf{r}}^{e}}|{{\phi }_{\beta }}({{\mathbf{k}}^{\prime }},t)\rangle =(1-{{\delta }_{\alpha \beta }})\delta (\mathbf{k}-{{\mathbf{k}}^{\prime }}){{\xi }_{\alpha \beta }},
\end{aligned}
\label{2req}
\end{equation}
where we have revived the momentum index $\mathbf{k}$ and 
$
{{\xi }_{\alpha \beta }}=\langle {{\phi }_{\alpha }}(\mathbf{k},t)|i{{\partial }_{\mathbf{k}}}|{{\phi }_{\beta }}(\mathbf{k},t)\rangle
$
is the generalized Berry connection. It is obvious from Eq.~(\ref{2req}) that the intraband components are not well-behaved and demonstrate singularities. 
To address the singularity issue \cite{sipe1995}, we decompose the matrix elements of the position operator's commutator into intraband and interband components. We extend this method to Floquet systems as follows:
\begin{equation}
\begin{aligned}
  & \langle {{\phi }_{\alpha }}(t)|[\mathbf{r},\rho ]|{{\phi }_{\beta }}(t)\rangle ={{[{{\mathbf{r}}^{i}},\rho ]}_{\alpha \beta }}+{{[{{\mathbf{r}}^{e}},\rho ]}_{\alpha \beta }}, \\ 
 & {{[{{\mathbf{r}}^{e}},\rho ]}_{\alpha \beta }}=\sum\nolimits_{\gamma }{\mathbf{r}_{\alpha \gamma }^{e}{{\rho }_{\gamma \beta }}-{{\rho }_{\alpha \gamma }}\mathbf{r}_{\gamma \beta }^{e}}, \\ 
 & {{[{{\mathbf{r}}^{i}},\rho ]}_{\alpha \beta }}=i{{\left( {{\rho }_{\alpha \beta }} \right)}_{;\mathbf{k}}} =i{{\partial }_{\mathbf{k}}}{{\rho }_{\alpha \beta }}+{{\rho }_{\alpha \beta }}({{\xi }_{\alpha \alpha }}-{{\xi }_{\beta \beta }}). \\ 
\end{aligned}
\label{rmatrix}
\end{equation}

\begin{figure*}
\includegraphics[width=\linewidth,trim={0 0 0.25cm 0}, clip]{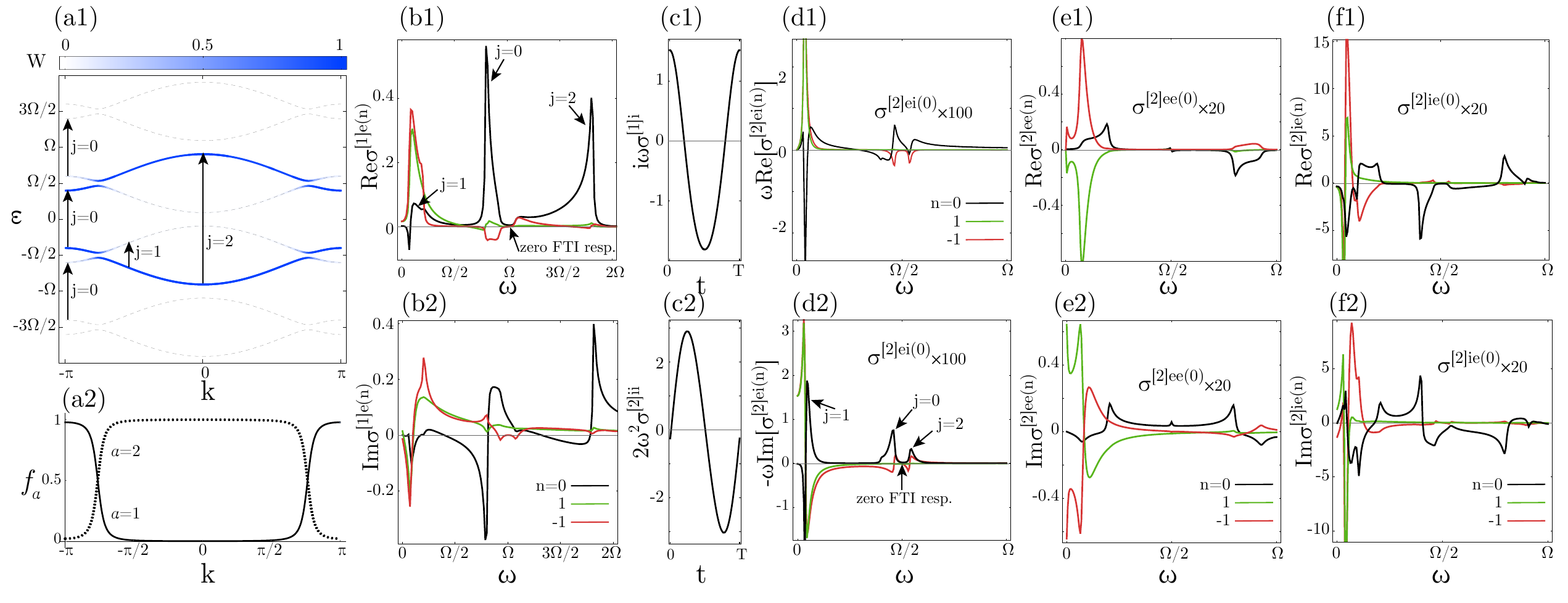} 
\caption{(Color online) Quasienergy band structure (a1), quench occupation of Floquet bands (a2), linear optical conductivity (b,c1) and nonlinear optical conductivity (c2,d,e,f) of the driven one-dimensional model defined in Eq.~(\ref{1ddriven}). The color scale in (a1) shows the physical weights $W^n_\alpha$ and $ j$-photon-assisted optical transitions are marked with arrows. The parameters used are $\omega_1=\omega_2=\omega$ and $A=0.3$. The linear (nonlinear) conductivities are expressed in units of $e^2/\hbar (e^3/\hbar)$ and $\omega$ is in units of $\Omega$. The $\sigma^{[2](0)}$ in panels (d-f) is scaled which are shown on the plots.}
\label{bands}
\end{figure*}
\subsection{First order conductivity}
Now, we develop the first-order perturbation theory. We define the perturbation as:  $V(t)=E z{{e}^{-i{{\omega }_{1}}t}}$ where $z=\mathbf{r}\cdot\mathbf{E}/E=z^e+z^i$ in Eq.~(\ref{rho1int}), i.e. the position operator's contribution can be separated into interband and intraband components. This separation leads to $\rho^{[1]}=\rho^{[1]e}+\rho^{[1]i}$, which can be evaluated using
Eqs.~(\ref{rho1int}) and (\ref{rmatrix}) (see Appendix at \cite{Note2}).

We find the expectation value of current $\langle \mathbf{J}\rangle =-\mathcal{N}\langle {{\mathbf{v}}}\rangle $ (setting the number of charges in the volume $\mathcal{N}=1$ hereafter and ${{v}^{x}}$ stands for velocity operator along the probe) and assume an expansion $\mathbf{J}={{\mathbf{J}}^{[0]}}+\lambda {{\mathbf{J}}^{[1]}}+{{\lambda }^{2}}{{\mathbf{J}}^{[2]}}+...$ and define $\langle {{\mathbf{J}}^{[1]}}\rangle =  {{\sigma }^{[1]}}E{{e}^{-i{{\omega }_{1}}t}} $, so ${{\sigma }^{[1]}}=-\text{Tr}({{v}^{x}}{{\rho }^{[1]}})/({Ee^{-i\omega_1t}})$.  The calculation shows that at the first order of perturbation when a probe electric field with a frequency of ${{\omega }_{1}}$ is applied to the system, the response would be at a frequency of ${{\omega }_{1}}+n\Omega,\,\,n\in \mathbb{Z}$ so ${{\sigma }^{[1]}}={{\sum\limits_{n}{e}}^{-in\Omega t}}{{\sigma }^{[1](n)}}$. As shown in \cite{Note2} we find
\begin{eqnarray}
\sigma _{xz}^{[1]e(n)}(\omega_1)&&=\sum\limits_{j\alpha \beta \mathbf{k}}{{{f}_{\beta \alpha }}\frac{v_{\beta \alpha }^{x(j+n)}z_{\alpha \beta }^{e(-j)}}{{{\epsilon }_{\alpha \beta }}+j\Omega -{{\omega }_{1}}}}
\label{sig1en}\\
{{\sigma }^{[1]i(n)}_{xz}}(\omega_1)&&=\sum\limits_{\alpha \mathbf{k}}{v_{\alpha \alpha }^{x(n)}\frac{1}{i{{\omega }_{1}}}{{\partial }_{{{k}_z}}}{{f}_{\alpha }}},
\label{sig1in}
\end{eqnarray}
where the sum over momentum $\mathbf{k}$ indicates the integral over the Brillouin zone $\sum_\mathbf{k}=\int d^n\mathbf{k}/(2\pi)^n$.
It is beneficial to express the matrix elements of the position operator in terms of the matrix elements of the velocity operator \cite{Note2} as:
\begin{equation}
\begin{aligned}
 \mathbf{r}_{\alpha \ne \beta }^{e(-j)}=\frac{-i}{{{\epsilon }_{\alpha \beta }}+j\Omega} \mathbf{v}_{\alpha  \beta }^{(-j)},
\end{aligned}
\label{rjvj}
\end{equation}
where ${\bf v}_{\beta\alpha  }^{(j)}=1/T\int _{0}^{T}{{e}^{ij\Omega t}}\langle {{\phi }_{\beta}}(t)|\partial_{\bf k} H({\bf k},t)|{{\phi }_{\alpha  }}(t)\rangle dt$. This is advantageous for numerical calculations as it eliminates the need to compute the derivative of Floquet quasi-modes, which would otherwise require defining a smooth gauge over the Brillouin zone.

Equation~(\ref{sig1en}) can be written in terms of only velocity matrix elements using 
Eq.~(\ref{rjvj}) and is consistent with the results previously obtained in \cite{seradjeh2020,deh2015,dabiri3}.  Equation~(\ref{sig1en}) for $n=0$ is very similar to the result of static systems with only a difference in a substitution 
$
 {{\epsilon }_{\alpha \beta }}\to {{\epsilon }_{\alpha \beta }}+j\Omega $ and  
 ${\mathbf{v}_{\beta \alpha }}\to \mathbf{v}_{\beta \alpha }^{(j)}
$ and its interpretation is quite simple in terms of optical transitions. Figure~\ref{bands}(a1) illustrates the different quasienergy bands defined in extended space. The undriven Hamiltonian has two bands, denoted by $\alpha = 1,2$. We define the optical transition associated with $j$ in Eq.~(\ref{sig1en}) as a "j-photon assisted" optical transition.
In Fig.~\ref{bands}(a1), we represent the optical transitions for different values of $j$ (0, 1, and 2) between the $\alpha = 1$ and $\alpha = 2$ bands. The total linear response is obtained by summing over all $j$ values and different initial sidebands.

Based on the interpretation outlined above, Eq.~(\ref{sig1en}) reveals that $\sigma_{x\neq z}^{[1]e(0)}$ represents the sum of Hall conductivities for occupied bands across all sidebands. For ideal occupation, this sum is quantized according to the Thouless-Kohmoto-Nightingale-den Nijs formula \cite{deh2015,seradjeh2020,dabiri3}. It is proportional to the Chern number of occupied bands, $\mathcal{C}_{occ}$, as:
$\sigma_{x\neq z}^{[1]e(0)} = \frac{1}{2\pi}\sum\mathcal{C}_{occ}$.

\subsection{Second order conductivity}
We calculate the second-order conductivity by obtaining the second-order density matrix. We assume $V(t)=E_2y{{e}^{-i{{\omega }_{2}}t}}$ and since $y$ and $\rho^{(1)}$ encompass the interband and intraband components as $y=y^i+y^e$ and $\rho^{[1]}=\rho^{[1]i}+\rho^{[1]e}$, subsequently, there will be four terms corresponding to $y^e,\rho^{[1]e}$ (interband-interband), $y^e,\rho^{[1]i}$ (interband-intraband), $y^i,\rho^{[1]e}$ (intraband-interband) and $y^i,\rho^{[1]i}$ (intraband-intraband) terms. Each term can be calculated separately using Eqs.~(\ref{rho1int}), and (\ref{rmatrix}) to find the $\rho^{[2]}$ as shown in \cite{Note2}.

 When two probe electric fields with frequencies of ${{\omega }_{1}},{{\omega }_{2}}$ are applied to the system, the second order response would be at a frequency of ${{\omega }_{1}}+{{\omega }_{2}}+n\Omega,\,\,n\in \mathbb{Z}$, therefore, ${{\sigma }^{[2]}}={{\sum\limits_{n}{e}}^{-in\Omega t}}{{\sigma }^{[2](n)}}$. Defining $\langle {{\mathbf{J}}^{[2]}}\rangle =  {{\sigma }^{[2]}}E E_2{{e}^{-i({{\omega }_{1}}+\omega_2)t}}$, we will find an expression for second-order conductivity as: \cite{Note2}

 \begin{eqnarray}
\begin{aligned}
& \sigma _{xyz}^{[2]ee(n)}=-\frac{1}{2}\sum\limits_{\alpha \beta \gamma {{j}_{1}}{{j}_{2}}\mathbf{k}}{\frac{v_{\beta \alpha }^{x({{j}_{1}}+{{j}_{2}}+n)}}{{{\epsilon }_{\alpha \beta }}-{{\omega }_{2}}-{{\omega }_{1}}+({{j}_{1}}+{{j}_{2}})\Omega }} \\ 
 & ~~~~~~~~~~~~~\times \{\frac{{{f}_{\beta \gamma }}y_{\alpha \gamma }^{e(-{{j}_{2}})}z_{\gamma \beta }^{e(-{{j}_{1}})}}{{{\epsilon }_{\gamma \beta }}-{{\omega }_{1}}+{{j}_{1}}\Omega }-\frac{{{f}_{\gamma \alpha }}y_{\gamma \beta }^{e(-{{j}_{1}})}z_{\alpha \gamma }^{e(-{{j}_{2}})}}{{{\epsilon }_{\alpha \gamma }}-{{\omega }_{1}}+{{j}_{2}}\Omega }\} \\ 
 & ~~~~~~~~~~~~~+\left( ({{\omega }_{1}},y)\leftrightarrow ({{\omega }_{2}},z) \right)
\\
& \sigma _{xyz}^{[2]ei(n)}=\frac{i}{2}\sum\limits_{\alpha \beta j \mathbf{k}}{\frac{v_{\beta \alpha }^{x(j+n)}y_{\alpha \beta }^{e(-j)}{{\partial }_{{{k}_{z}}}}{{f}_{\beta \alpha }}}{{{\omega }_{1}}({{\epsilon }_{\alpha \beta }}+j\Omega -{{\omega }_{1}}-{{\omega }_{2}})}} \\ 
& ~~~~~~~~~~~~~+\left( ({{\omega }_{1}},y)\leftrightarrow ({{\omega }_{2}},z) \right)
\\
& {{\sigma }_{xyz}^{[2]ie(n)}}=\sum\limits_{\alpha \beta j \mathbf{k}}{\frac{-iv_{\beta \alpha }^{x(j+n)}/2}{{{\epsilon }_{\alpha \beta }}-{{\omega }_{1}}-{{\omega }_{2}}+j\Omega }{{\partial }_{{{k}_{y}}}}\left( \frac{{{f}_{\beta \alpha }}z_{\alpha \beta }^{e(-j)}}{{{\epsilon }_{\alpha \beta }}+j\Omega -{{\omega }_{1}}} \right)} \\ 
 & -\frac{1}{2}\sum\limits_{\alpha \beta j{{j}_{2}}\mathbf{k}}{\frac{v_{\beta \alpha }^{x(j+{{j}_{2}}+n)}{{f}_{\beta \alpha }}z_{\alpha \beta }^{e(-j)}}{{{\epsilon }_{\alpha \beta }}-{{\omega }_{1}}-{{\omega }_{2}}+(j+{{j}_{2}})\Omega }\frac{(\xi _{\alpha \alpha }^{y(-{{j}_{2}})}-\xi _{\beta \beta }^{y(-{{j}_{2}})})}{{{\epsilon }_{\alpha \beta }}+j\Omega -{{\omega }_{1}}}} \\ 
 & ~~~~~~~~~~~~+(({{\omega }_{1}},z)\leftrightarrow ({{\omega }_{2}},y))
\\
&\sigma _{xyz}^{[2]ii(n)}=\frac{1}{2}\sum\limits_{\alpha \mathbf{k}}{\frac{v_{\alpha \alpha }^{x(n)}{{\partial }_{{{k}_{z}}}}{{\partial }_{{{k}_{y}}}}{{f}_{\alpha }}}{{{\omega }_{1}}({{\omega }_{1}}+{{\omega }_{2}})}}+{{\omega }_{1}}\leftrightarrow {{\omega }_{2}}
\label{sigmaha}
  \end{aligned}
\end{eqnarray}

 The second-order response of Floquet systems at a frequency of $\omega_1 + \omega_2$ resembles that of static systems, with one key difference: the optical transitions are replaced by $j_1$- and $j_2$-photon-assisted transitions. Equation~(\ref{sigmaha}) is recast to the formulas for the nonlinear response of static systems \cite{watan2021} in the limit of time-independent Hamiltonian. Notice that these frequency-dependent features of the linear and nonlinear optical conductivity in Floquet systems are intricately linked to the driving frequency, influencing band structure, topological properties, and the balance between different optical processes. 

{For the sake of completeness, we add a section in the Appendix showing the formulas for the third order nonlinear response of Floquet systems. Indeed according to the recursive relation given by Eq.~(\ref{rho1int}) and given the density matrix of order $n$ i.e. $\rho^{[n]}$ it is straightforward to calculate the $(n+1)$ order response.}

\subsection{Special limits}
We will consider various scenarios. 
i) AC response to a DC electric field:
Assume that $\omega_1=\omega_2=0$, and hence the full-intraband linear and second-order conductivities diverge based on Eqs.~(\ref{sig1in}) and (\ref{sigmaha}). This is a real divergence (not a spurious one usually arising in a velocity gauge \cite{taghi2017}) and we call it \emph{Drude peak at finite frequency}, since the responses in Eqs.~(\ref{sig1in}) and (\ref{sigmaha}) are at frequencies $n\Omega,~\,n\in \mathbb{Z}$. This process, which is essentially the reverse of rectification, is experimentally intriguing and could have significant implications for the field of optoelectronics. {The importance of the Drude peak lies in its ability to characterize material properties, understand charge transport phenomena, and predict system behavior in a wide range of scientific and technological applications. It is a rich source of information about a material's electrical, optical, and transport properties. Its analysis is fundamental in both understanding basic physical phenomena and advancing technologies. Having engineered the occupation of Floquet states, which is a function of switching protocol, temperature, coupling to the bath and other dissipation mechanisms, this response could be modulated at will which may be useful to develop special optical devices.} 

ii) Zero response of Floquet topological insulators (FTIs):
We refer to FTIs as the Floquet systems having finite gaps at quasienergies $n \Omega/2,~n\in\mathbb{Z}$. Then at frequencies $\omega_1=n\Omega$, there are no resonant optical transitions as shown in Fig.~\ref{bands}(a1). Let us consider for instance the first order conductivity $\sigma^{[1]e}_{xx}$ in Eq.~(\ref{sig1en}). Using Eq.~(\ref{rjvj}) and adding an infinitesimal constant $\eta$ to frequency $\omega_1$ to account for the relaxation, it can be written as
\begin{equation}
\begin{aligned}
\operatorname{Re}\sigma _{xx}^{[1]e(0)}=\sum\limits_{j\alpha \beta }{\pi\frac{{{f}_{\beta \alpha }}|v_{\beta \alpha }^{x(j)}{{|}^{2}}\delta ({{\epsilon }_{\alpha \beta }}+j\Omega -{{\omega }_{1}})}{({{\epsilon }_{\alpha \beta }}+j\Omega )}},
\end{aligned}
\label{sig1exx}
\end{equation}
where we have used the approximation $\frac{1}{x-i\eta}\approx \frac{x}{x^2+\eta^2}+i\pi \delta(x)$ with $\delta(x)$ the Dirac delta function. Since for FTIs and $\omega_1=n\Omega$ the delta function in Eq.~(\ref{sig1exx}) is always zero, then $\operatorname{Re}\sigma _{xx}^{[1]e(0)}$ would be zero as well. This is in agreement with previous results \cite{seradjeh2020,wu2011}, albeit this effect may be suppressed when gaps are small and the broadening factor $\eta$ is large.

{iii) High harmonic generation: Interestingly, arbitrary-order harmonic generation can be achieved in Floquet systems at the first order of perturbation theory, as shown in Eq.~(\ref{sig1in}), by setting $\omega = \Omega$. This contrasts with static systems, where $n$th-order harmonic generation requires at least the $n$th order of perturbation theory. According to our formalism, we predict a phenomenon specific to two dimensions: the nonlinear Hall effect at integer multiples of the drive frequency. When a current is applied to the driven system, a Hall voltage is generated at any integer multiple of the drive frequency.}

iv) DC photocurrent:
The DC response to an AC electric field is usually called photocurrent. It is obvious from Eqs.~(\ref{sig1en}) and (\ref{sig1in}) that when an electric field at frequency of $m\Omega,~m \in \mathbb{Z}$ is applied to the system, then $\sigma^{[1]e(-m)}$ and $\sigma^{[1]i(-m)}$ give DC photocurrents. This is in contrast to static systems where there is no photocurrent at the first order of perturbation theory. 

At the second order of perturbation, when $\omega_1+\omega_2=0$, there are DC photocurrents and all previous formulas of photocurrents \cite{watan2021} including shift current, injection current, Berry dipole contribution, etc. are generalized~\cite{Note2} to Floquet systems $\sigma^{[2]}\rightarrow \sigma^{[2](0)}$ which is similar to formulas for multiband static systems.

When $\omega_1+\omega_2=m\Omega,~m \in \mathbb{Z}$, there are also DC photocurrents which are obtained by $\sigma^{[2](-m)}$ in Eq.~(\ref{sigmaha}). This means that the probe electric field can make resonance with the Floquet system giving rise to DC photocurrent.
It should be noted that the higher orders of nonlinear response can be straightforwardly derived using Eqs.~(\ref{rho1int}) and (\ref{rmatrix}).

\begin{figure*}
\includegraphics[width=\linewidth,trim={0 0 0.25cm 0}, clip]{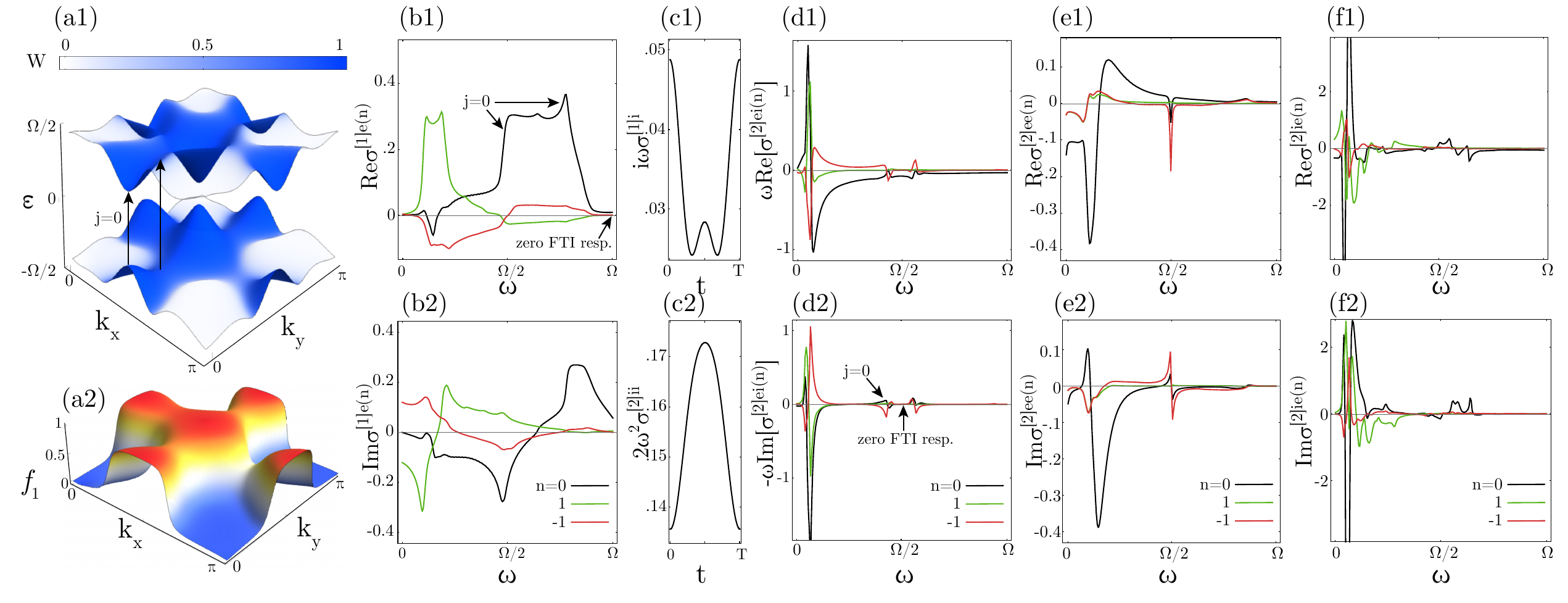} 
\caption{ Quasienergy band structure (a1), quench occupation of first Floquet band (a2), linear optical conductivity (b,c1), and nonlinear optical conductivity (cc-f) of the driven 2D quantum well model defined in Eq.~(\ref{floquetwell}). The color scale in (a1) shows the physical weights $W^n_\alpha$ and $ j$-photon-assisted optical transitions are marked with arrows. The parameters are $\omega_1=\omega_2=\omega$ and $M=0.2, B=0.2, \Omega=2.5, A=0.35$. The linear (nonlinear) conductivities are in units of $e^2/\hbar (e^3/\hbar)$ and $\omega$ is in units of $\Omega$.}
\label{bands2D}
\end{figure*}

\section{Numerical Results}
Now, we turn our attention to implementing our approach by focusing on systems confined to one- and two-dimensional geometries. These lower-dimensional systems provide unique opportunities to explore phenomena that are otherwise absent or less pronounced in three-dimensional counterparts.

\subsection{One-dimensional system} 
Here we apply our findings for a driven one-dimensional two-band system. We take as a prototype the Su-Schrieffer-Heeger model proposed for electronic states in polyacetylene \cite{sshref} subjected to an oscillating field. The Hamiltonian in $\mathbf{k}$ space reads as:
\begin{equation}
\begin{aligned}
H(k,t)={{\sigma }_{x}}\left( 2+\cos k \right)+ {{\sigma }_{y}}\sin k+{{\sigma }_{z}}\left( 1+A\cos (\Omega t) \right),
\end{aligned}
\label{1ddriven}
\end{equation}
where $\sigma_i,~~i\in{x,y,z}$ are Pauli matrices. The quasienergy bands are shown for $A=0.3$ in Fig.~\ref{bands}(a1) where the color scale indicates the physical weights of subbands defined as $W^n_\alpha=\langle \phi_\alpha^{(n)}|\phi_\alpha^{(n)}\rangle \le 1 $. For Floquet systems the occupation of Floquet states is generally not easily described by Fermi-Dirac distribution and we assume a \emph{quench occupation} \cite{wu2011,seradjeh2020,dabiri3,PhysRevB.109.115431} of states which is obtained by projecting the non-driven ground states $|g_0\rangle$ on the Floquet states ${{f}_{\alpha }}=\sum\nolimits_{n}{|\langle {{g}_{0}}|\phi _{\alpha }^{(n)}\rangle {{|}^{2}}}$ and describes the situation where the drive is suddenly turned on. The occupations are shown in Fig.~\ref{bands}(a2) where there are band inversions at two resonances in the Brillouin zone.

Figures~\ref{bands}(b1) and (b2) depict the real and imaginary parts of linear interband conductivities according to Eq.~(\ref{sig1en}). The peaks of conductivities in Fig. \ref{bands}(b1) are associated with the optical transition between bands with higher physical weights and density of states which are marked by arrows emphasizing $ j$-photon-assisted transition origin.

The interesting response which was not highlighted in previous works is the divergent full intraband responses shown in Figs.~\ref{bands}(c1) and (c2) as a function of time which have DC and AC components and are in the same phase.

Figures~\ref{bands}(d1) and (d2) show the $\sigma^{[2]ei}$ which shares some similarities to linear interband response. According to Eq.~(\ref{sigmaha}) it contains the derivative of distribution and shows a divergent behavior at zero frequency. In Fig.~\ref{bands}(d2) we mark different peaks with arrows showing the relevant $j$-photon assisted transitions which can be traced back to Fig.~\ref{bands}(a1). They show a sharp peak at zero frequency and a long tail at higher frequencies for coherent Drude carriers. We also observe another peak near the $\Omega/2$, since the derivative of occupations has finite value only near the resonances (see Fig.~\ref{bands}(a2)).

The full interband conductivity is shown in Figs.~\ref{bands}(e1),(e2). The peaks are attributed to van Hove singularities in the density of states, and also the group velocity, which makes a shift in the frequencies of the peaks. Seemingly, in the middle frequency, the interband and intraband contributions to nonlinear optical conductivity have opposite effects as the frequency increases.

To calculate $\sigma^{[2]ie}$ we should find a smooth gauge for Floquet quasi modes because otherwise the derivatives in the third equation of Eq.~(\ref{sigmaha}) will not be well-defined. The result is depicted in Figs.~\ref{bands}(f1) and (f2) showing more peaks and dips with respect to  $\sigma^{[2]ee}, \sigma^{[2]ei}$ because the $2\omega$ resonances are enabled. High heterodyne $\sigma^{(\pm1)}$ responses to homodyne response $\sigma^{(0)}$ at low frequencies are notable in Figs.~\ref{bands}(d-f) which originates from the transitions near the dynamical gap at quasienergy $\epsilon=\pm\Omega/2$.

Although we assume quench occupation of Floquet states in Fig.~\ref{bands}, other occupations are possible such as the \emph{ideal occupation} \cite{deh2015,seradjeh2020,dabiri3} of states where the lower (upper) Floquet band is full (empty) and for which the derivative of occupation is zero so $\sigma^{[1]i}, \sigma^{[2]ei},\sigma^{[2]ii}$ are zero. This case is also studied and results are presented in \cite{Note2}. On the other hand for \emph{mean-energy} occupation \cite{wu2011,dabiri3} $f_ {\alpha}=\theta(-\bar{\epsilon}_\alpha)$ where $\bar{\epsilon}_\alpha=\epsilon_\alpha+\sum_n n\Omega W^n_{\alpha}$, the derivative of occupation to momentum is singular at resonances. The real occupation of states in experimental setups depends on the switch-on protocol and also dissipation mechanisms. There are also proposals for on-demand control of occupation of Floquet states using multifrequency drives and engineered switch-on protocol \cite{optimal2022} or coupling to Fermi and Bose baths \cite{seeth}.

\subsection{Two-dimensional system}
{We apply the formalism to a two-dimensional system to calculate the linear and nonlinear optical responses of a driven quantum well. The Hamiltonian of a quantum well, formed in certain semiconductor heterojunctions \cite{bhzpaper,seradjeh2020}, incorporating an inversion symmetry-breaking term and an oscillating field, can be expressed in $\mathbf{k}$-space as} 
\begin{eqnarray}
&&H\left( {{k}_{x}},{{k}_{y}},t \right)= \left[ 0.2+\sin \left( {{k}_{x}} \right) \right]{{\sigma }_{x}}+\sin ({{k}_{y}}){{\sigma }_{y}}\nonumber\\
 &&+\left[ M-4B+2B\cos \left( {{k}_{x}} \right)+2B\cos \left( {{k}_{y}} \right)+A\cos \left( \Omega \text{t} \right) \right]{{\sigma }_{z}}\nonumber\\
\label{floquetwell}
\end{eqnarray}
{The parameters $M,B$ determine the topological phase of the quantum well, $0.2 \sigma_z$ is the inversion symmetry breaking term and $A \cos(\Omega t) \sigma_z$ is due to the oscillating potential.}

{The quasi-energy band structure of the model (\ref{floquetwell}) and the quench occupation of Floquet states are represented in Figs.~\ref{bands2D}(a1) and (a2), respectively. It is seen that the quench occupation of states inverts at resonance (band inversion) points in the Brillouin zone. The linear and nonlinear optical conductivities are also shown in Figs.~\ref{bands2D}(b) and (f) assuming quench occupation of states. The peaks and dips in Figs.~\ref{bands2D}(b1) and (d2) are marked with arrows showing the corresponding $j$ photon-assisted optical transitions. Similar to the 1D system, the nonlinear components $\sigma^{[2]ee}$ and $\sigma^{[2]ei}$ only involve $\omega$ and $2\omega$-resonances, respectively while $\sigma^{[2]ie}$ includes both $\omega$ and $2\omega$-resonances thus showing more peaks and dips in its structure. The strong responses at low frequencies are due to transitions across the dynamical gap at $\varepsilon=\Omega/2$ and are an interesting feature of on-resonant drives with frequencies lower than bandwidth.}

\section{Conclusion} 
In this work, we developed a comprehensive theoretical framework for calculating the linear and nonlinear optical responses of Floquet systems within the length gauge formalism. By utilizing density matrix evolution in the Floquet basis, our approach captures both interband and intraband contributions, providing a robust method for analyzing dynamical optical responses across a wide range of systems and perturbation orders. { This framework not only generalizes existing theories but also reveals different phenomena, such as divergent AC responses to DC fields and j-photon-assisted optical transitions, which are intrinsically tied to the periodic driving of quantum systems.}

{Through numerical investigations of one- and two-dimensional systems, such as a quantum well and a driven Su-Schrieffer-Heeger model, we illustrated the adaptability of our framework. These examples demonstrated the special interaction between periodic driving and nonlinear optical effects, illuminating rich phenomena such as band inversions, Floquet band topology, and resonant optical transitions. The promise of Floquet systems for investigating and utilizing sophisticated quantum phenomena is further highlighted by the detection of characteristics such as high harmonic production, nonlinear Hall effects, and DC photocurrents.}

{Our results have important implications for both fundamental research and practical applications. In terms of theory, this work bridges the gap between driven and static systems by offering a single framework for investigating nonlinear optical responses in time-periodic quantum systems. Our paradigm provides fresh insights into how topology, band structure, and symmetry shape the optical properties of materials by providing a link between the dynamics of Floquet quasi-states and measurable quantities such as optical conductivity and photocurrents.}

{Practically speaking, periodic driving creates new opportunities for device innovation by enabling the engineering of optical responses. Applications include quantum information systems, sensors, ultrafast spintronic devices, and optoelectronic components. The versatility of this framework for experimental research and technology development is demonstrated by the capacity to manage phenomena like divergent AC responses or customised photocurrents by driving frequency, field strength, or occupancy procedures.}

{In conclusion, our theoretical approach contributes to the knowledge of dynamical optical responses in Floquet systems and establishes these systems as a promising platform for the exploration and utilization of quantum phenomena in driven quantum materials.}

\section{Acknowledgement}
We thank Tami Pereg-Barnea for fruitful discussions. R. A. received partial funding from the Iran National Science Foundation (INSF) under Project No. 4026871. S. S. D. thanks the late Seyed Mahdi Dabiri for his valuable help and support.
\\
\appendix

\section{Details of derivations}
This appendix provides detailed calculations referenced in the main text. We present the conductivities in the two-band model limit, along with the conductivities expressed in terms of velocity matrix elements. Additionally, we include a discussion of the covariant derivative and derive the DC photo-conductivities, incorporating the Berry curvature dipole term, injection current, and shift current for Floquet systems. We present the results of the optical conductivities for a one-dimensional driven model with the ideal occupation of Floquet states. { Finally, we find the formulas for optical conductivity of Floquet systems at the third order of perturbation theory.}

Floquet theorem is an essential theorem in describing the time-periodic systems. The Schr\" {o}dinger equation is recast for Floquet quasi modes as:
\begin{equation}
\begin{aligned}
(H(t)-i{{\partial }_{t}})|{{\phi }_{\alpha }}(t)\rangle ={{\epsilon }_{\alpha }}|{{\phi }_{\alpha }}(t)\rangle .
\end{aligned}
\label{schro2}
\end{equation}
Since the Hamiltonian and also quasi modes are time-periodic, we can Fourier expand them as ${{H}}(t)={{e}^{-in\Omega t}}{{H}^{(n)}}$ and $|{\phi }_{\alpha }(t)\rangle ={{e}^{-im\Omega t}}|{\phi }_{\alpha }^{(m)}\rangle $. By inserting the Fourier series in (\ref{schro2}) we find 
\begin{equation}
\begin{aligned}
{{\sum }_{n}}{{H}^{(m-n)}}|\phi _{\alpha }^{(n)}\rangle -m\Omega |\phi _{\alpha }^{(m)}\rangle ={{\epsilon }_{\alpha }}|\phi _{\alpha }^{(m)}\rangle .
\end{aligned}
\label{hmn}
\end{equation}
We can write the Schr\"{o}dinger equation i.e. Eq.~(\ref{hmn}) in a matrix form as $\mathcal{H}{{\varphi }_{\alpha }}={{\varepsilon }_{\alpha }}{{\varphi }_{\alpha }}$ where
\begin{equation}
\begin{aligned}
\mathcal{H}=\left( \begin{matrix}
   \ddots  & {} & {} & {} & {}  \\
   {} & {{H}^{(0)}}+\Omega  & {{H}^{(-1)}} & {{H}^{(-2)}} & {}  \\
   {} & {{H}^{(1)}} & {{H}^{(0)}} & {{H}^{(-1)}} & {}  \\
   {} & {{H}^{(2)}} & {{H}^{(1)}} & {{H}^{(0)}}-\Omega  & {}  \\
   {} & {} & {} & {} & \ddots   \\
\end{matrix} \right){{\varphi }_{\alpha }}=\left( \begin{matrix}
   \vdots   \\
   \phi _{\alpha }^{(-1)}  \\
   \phi _{\alpha }^{(0)}  \\
   \phi _{\alpha }^{(1)}  \\
   \vdots   \\
\end{matrix} \right)
\end{aligned}
\label{bigh}
\end{equation}
 On this basis, we effectively have a simple Schrödinger equation with a time-independent Hamiltonian, $\mathcal{H} \varphi_{\alpha} = \varepsilon_{\alpha} \varphi_{\alpha}$, in an extended space labeled by $\alpha = 1, 2, \dots, N_b$ and $n = -N_s, \dots, -1, 0, 1, \dots, N_s$. Thus, $\mathcal{H}$ encompasses a total of $2N_b (2N_s + 1)$ bands. We refer to bands distinguished by different $n$ indices as different sidebands.
It can be shown that in the large matrix limit \cite{kitagawa2011}, if $(..., \phi_{\alpha}^{(-1)}, \phi_{\alpha}^{(0)}, \phi_{\alpha}^{(1)}, \dots)^{T}$ is an eigenfunction of $\mathcal{H}$, then the shifted wave vector $(..., \phi_{\alpha}^{(-2)}, \phi_{\alpha}^{(-1)}, \phi_{\alpha}^{(0)}, \dots)^{T}$ is also an eigenfunction of $\mathcal{H}$. Since $\mathcal{H}$ is Hermitian, its wavefunctions can be taken to be orthonormal. Therefore, we can assume $\sum\nolimits_{m}{\langle \phi _{\alpha }^{(m)}|\phi _{\beta }^{(m-j)}\rangle }={{\delta }_{\alpha \beta }}{{\delta }_{j0}} $
where $\delta $ is the Kronecker delta. This condition can be written as:
\begin{equation}
\begin{aligned}
\langle {{\phi }_{\alpha }}(t)|{{\phi }_{\beta }}(t)\rangle ={{\delta }_{\alpha \beta }}.
\end{aligned}
\label{orthoeq}
\end{equation}
Equation~(\ref{orthoeq}) is the orthonormality condition that we assume hereafter. 

\subsection{perturbation theory}
We derive Eq.~(\ref{rho1int}) of the main text. Substituting the density matrix expansion in the von Neumann equation and equating the terms with the same powers of $\lambda $ leads to
\begin{equation}
\begin{aligned}
 & i{{\partial }_{t}}{{\rho }^{[n]}}=[H(t),{{\rho }^{[n]}}]+[V(t),{{\rho }^{[n-1]}}], \\ 
\end{aligned}
\label{3rho}
\end{equation}
where $n$ as studied here is 0, 1, 2, ... with ${\rho }^{[-1]}=0$. Using Eq.~(\ref{schro2}) one can write
\begin{eqnarray}
   \langle {{\phi }_{\alpha }}(t)|[H(t),{{\rho }^{[n]}}]|{{\phi }_{\beta }}(t)\rangle &&={{\epsilon }_{\alpha\beta }}\langle {{\phi }_{\alpha }}(t)|{{\rho }^{[n]}}|{{\phi }_{\beta }}(t)\rangle \nonumber\\
  &&-i\langle {{\partial }_{t}}{{\phi }_{\alpha }}(t)|{{\rho }^{[n]}}|{{\phi }_{\beta }}(t)\rangle \nonumber\\
  &&-i\langle {{\phi }_{\alpha }}(t)|{{\rho }^{[n]}}|{{\partial }_{t}}{{\phi }_{\beta }}(t)\rangle.   
\label{hrho1}
\end{eqnarray}
Inserting Eq.~(\ref{hrho1}) in Eq.~(\ref{3rho}) leads to
\begin{equation}
\begin{aligned}
i{{\partial }_{t}}\rho _{\alpha \beta }^{[n]}={{\epsilon }_{\alpha \beta }}\rho _{\alpha \beta }^{[n]}+\langle {{\phi }_{\alpha }}(t)|[V(t),{{\rho }^{[n-1]}}]|{{\phi }_{\beta }}(t)\rangle 
\end{aligned}
\label{rho1eq0}
\end{equation}
We can change the variable as $\rho _{\alpha \beta }^{[n]}=S(t){{e}^{-i{{\epsilon }_{\alpha \beta }}t}}$. So $i{{\partial }_{t}}\rho _{\alpha \beta }^{[n]}=i{{\partial }_{t}{S}}(t){{e}^{-i{{\epsilon }_{\alpha \beta }}t}}+{{\epsilon }_{\alpha \beta }}S(t){{e}^{-i{{\epsilon }_{\alpha \beta }}t}}$. Then the equation for $S(t)$ in (\ref{rho1eq0}) can be easily integrated to yield Eq.~(\ref{rho1int}).

\subsection{First order conductivity}

By defining $V(t)=Ez{{e}^{-i{{\omega }_{1}}t}}$ (where $z=\mathbf{r}\cdot\mathbf{E}/E$ is the position operator along the electric field) and using Eqs.~(\ref{rho1int}) and (\ref{rmatrix})) the interband density matrix can be written as:
\begin{equation}
\begin{aligned}
   \rho _{\alpha \beta }^{[1]e}&=-iE{{e}^{-i{{\epsilon }_{\alpha \beta }}t}}\int_{-\infty }^{t}{{{e}^{i{{\epsilon }_{\alpha \beta }}t'}}\langle {{\phi }_{\alpha }}({{t}^{\prime }})|[z^e e^{-i\omega_1t'},{{\rho }^{[0]}}]|{{\phi }_{\beta }}({{t}^{\prime }})\rangle d{{t}^{\prime }}}\\
&=-i{{e}^{-i{{\epsilon }_{\alpha \beta }}t}}E\sum\limits_{\gamma }{\int_{-\infty }^{t}{{{e}^{i({{\epsilon }_{\alpha \beta }}-{{\omega }_{1}})t'}}}}  \left( {{z}^e_{\alpha \gamma }}{{\rho }^{[0]}_{\gamma \beta }}-{{\rho }^{[0]}_{\alpha \gamma }}{{z}^e_{\gamma \beta }} \right)d{{t}^{\prime }} \\ 
 & =-i{{e}^{-i{{\epsilon }_{\alpha \beta }}t}}E\int_{-\infty }^{t}{{{e}^{i({{\epsilon }_{\alpha \beta }}-{{\omega }_{1}})t'}}({{f}_{\beta }}-{{f}_{\alpha }}){{z}^e_{\alpha \beta }}d{{t}^{\prime }}} \\ 
 & =-E\sum\limits_{mn}{\frac{{{e}^{i(-{{\omega }_{1}}+(m-n)\Omega )t}}}{{{\epsilon }_{\alpha \beta }}-{{\omega }_{1}}+(m-n)\Omega }({{f}_{\beta }}-{{f}_{\alpha }})\langle \phi _{\alpha }^{(m)}|z^e|\phi _{\beta }^{(n)}\rangle } \\ 
 & =-E\sum\limits_{j}{\frac{{{e}^{i(-{{\omega }_{1}}+j\Omega )t}}}{{{\epsilon }_{\alpha \beta }}+j\Omega -{{\omega }_{1}}}{{f}_{\beta \alpha }}{z}_{\alpha \beta }^{e(-j)}} \\ 
\end{aligned}
\label{rho1emat}
\end{equation}
where 
\begin{eqnarray}
z_{\alpha \beta }^{e(-j)}=&&\sum\nolimits_{m}{\langle \phi _{\alpha }^{(m)}|{{z}^{e}}|\phi _{\beta }^{(m-j)}\rangle }\nonumber\\
=&&1/T\int _{0}^{T}{{e}^{-ij\Omega t}}\langle {{\phi }_{\alpha }}(t)|{{z}^{e}}|{{\phi }_{\beta }}(t)\rangle dt.
\label{zab}
\end{eqnarray}
 Then we can find the first-order interband conductivity by evaluating the expectation value of the velocity operator as
 \begin{widetext}
\begin{equation}
\begin{aligned}
  & {{\sigma }^{[1]e}}=-\frac{\text{Tr}({{v}^{x}}{{\rho }^{[1]e}})}{E{{e}^{-i{{\omega }_{1}}t}}}=-\frac{1}{E{{e}^{-i{{\omega }_{1}}t}}}\sum\limits_{\alpha \beta \mathbf{k} }{v_{\beta \alpha }^{x}\rho _{\alpha \beta }^{[1]e}} =\sum\limits_{npsj\alpha \beta \mathbf{k} }{\langle \phi _{\beta }^{(n)}|{{v}^{x\{s\}}}|\phi _{\alpha }^{(p)}\rangle \frac{{{e}^{i(j+n-p-s)\Omega t}}}{{{\epsilon }_{\alpha \beta }}+j\Omega -{{\omega }_{1}}}{{f}_{\beta \alpha }}z_{\alpha \beta }^{(-j)}} \\
&\sigma _{xz}^{[1]e(n)}=\sum\limits_{j\alpha \beta \mathbf{k}}{{{f}_{\beta \alpha }}\frac{v_{\beta \alpha }^{x(j+n)}z_{\alpha \beta }^{e(-j)}}{{{\epsilon }_{\alpha \beta }}+j\Omega -{{\omega }_{1}}}}
\end{aligned}
\label{}
\end{equation}
where 
\begin{equation}
\begin{aligned}
v_{\beta\alpha  }^{x(j)}=\sum\nolimits_{m s}{\langle \phi _{\beta }^{(m)}|{{v}^{x\{s\}}}|\phi _{\alpha  }^{(m+j-s)}\rangle }=1/T\int _{0}^{T}{{e}^{ij\Omega t}}\langle {{\phi }_{\beta}}(t)|{{v}^x}|{{\phi }_{\alpha  }}(t)\rangle dt
\end{aligned}
\label{vab}
\end{equation}
\end{widetext}
 and since the velocity operator ${v}^{x}=\partial_{k_x} H(t)$ may be time-periodic with period $T$ we Fourier expand it as ${v}^{x}=\sum_s{{e}^{-is\Omega t}}v^{x\{s\}}$.
Also for the intraband density matrix one can write
\begin{eqnarray}
  \rho _{\alpha \beta }^{[1]i}=&&-iE\int_{-\infty }^{t}{{{e}^{-i{{\omega }_{1}}t'}}\langle {{\phi }_{\alpha }}({{t}^{\prime }})|[{{z}^{i}},{{\rho }^{[0]}}]|{{\phi }_{\beta }}({{t}^{\prime }})\rangle d{{t}^{\prime }}} \nonumber\\ 
  =&&-iE{{\delta }_{\alpha \beta }}\int_{-\infty }^{t}{{{e}^{-i{{\omega }_{1}}t'}}i{{\partial }_{{{k}_{z}}}}\rho _{\alpha \alpha }^{[0]}}  \nonumber\\
  =&&E{{\delta }_{\alpha \beta }}\int_{-\infty }^{t}{{{e}^{-i{{\omega }_{1}}t'}}{{\partial }_{{{k}_{z}}}}{{f}_{\alpha }}}=\frac{E{{e}^{-i{{\omega }_{1}}t}}}{-i{{\omega }_{1}}}{{\delta }_{\alpha \beta }}{{\partial }_{{{k}_{z}}}}{{f}_{\alpha }}\nonumber\\  
\label{rho1i}
\end{eqnarray}
So
\begin{equation}
\begin{aligned}
  & {{\sigma }_{xz}^{[1]i}}=-\frac{\text{Tr}({{v}^{x}}{{\rho }^{[1]i}})}{E{{e}^{-i{{\omega }_{1}}t}}}=-\frac{\sum\limits_{\alpha }{v_{\alpha \alpha }^{x}\rho _{\alpha \alpha }^{[1]i}}} {E{{e}^{-i{{\omega }_{1}}t}}}=\sum\limits_{\alpha }{v_{\alpha \alpha }^{x}\frac{1}{i{{\omega }_{1}}}{{\partial }_{{{k}_z}}}{{f}_{\alpha }}} \\ 
\end{aligned}
\label{sig1i}
\end{equation}

\subsection{Second order conductivity}
Now we find the second order conductivity formulas using Eqs.~(\ref{rho1int}) and (\ref{rmatrix}) and first order density matrix which was derived in the previous section. According to Eq.~(\ref{rho1int})  one can write
\begin{equation}
\begin{aligned}
{\rho }^{[2]}_{\alpha \beta }=-i{{e}^{-i{{\epsilon }_{\alpha \beta }}t}}\int_{-\infty }^{t}{{{e}^{i{{\epsilon }_{\alpha \beta }}t'}}\langle {{\phi }_{\alpha }}({{t}^{\prime }})|[V({{t}^{\prime }}),{{\rho }^{[1]}}]|{{\phi }_{\beta }}({{t}^{\prime }})\rangle d{{t}^{\prime }}}.
\end{aligned}
\label{rho2int}
\end{equation}
We should find the matrix elements in the above equation as:
\begin{equation}
\begin{aligned}
  & \langle {{\phi }_{\alpha }}({{t}^{\prime }})|[V(t'),{{\rho }^{[1]}}]|{{\phi }_{\beta }}({{t}^{\prime }})\rangle ={{E}_{2}}{{e}^{-i{{\omega }_{2}}t'}}\langle {{\phi }_{\alpha }}({{t}^{\prime }})|[y,{{\rho }^{[1]}}]|{{\phi }_{\beta }}({{t}^{\prime }})\rangle  \\ 
\end{aligned}
\label{comyr1}
\end{equation}
Since $y$ and $\rho^{[1]}$ include the interband and intraband components as $y=y^i+y^e$ and  $\rho^{[1]}=\rho^{[1]i}+\rho^{[1]e}$, in the above equation there are four terms corresponding to $y^e,\rho^{[1]e}$ (interband-interband), $y^e,\rho^{[1]i}$ (interband-intraband), $y^i,\rho^{[1]e}$ (intraband-interband) and $y^i,\rho^{[1]i}$ (intraband-intraband) terms. We calculate each term separately in the following sections.

\subsubsection{Interband-interband conductivity}
Using Eq.~(\ref{rmatrix}) we can write
\begin{equation}
\begin{aligned}
\langle {{\phi }_{\alpha }}({{t}^{\prime }})|[{{y}^{e}},{{\rho }^{[1]e}}]|{{\phi }_{\beta }}({{t}^{\prime }})\rangle =\sum\limits_{\gamma }{\{y_{\alpha \gamma }^{e}\rho _{\gamma \beta }^{[1]e}-\rho _{\alpha \gamma }^{[1]e}y_{\gamma \beta }^{e}\}}
\end{aligned}
\label{}
\end{equation}
Substituting (\ref{rho1emat}) in the above equation gives
\begin{equation}
\begin{aligned}
   & {{E}_{2}}{{e}^{-i{{\omega }_{2}}t'}}\langle {{\phi }_{\alpha }}({{t}^{\prime }})|[{{y}^{e}},{{\rho }^{[1]e}}]|{{\phi }_{\beta }}({{t}^{\prime }})\rangle  \\ 
 & =-E{{E}_{2}}\sum\limits_{{{j}_{1}\gamma}}{{{e}^{i(-{{\omega }_{1}}-{{\omega }_{2}}+{{j}_{1}}\Omega )t'}}}\\
&~~~ \times \left[ \frac{{{f}_{\beta \gamma }}{{y}^e_{\alpha \gamma }}z_{\gamma \beta }^{e(-{{j}_{1}})}}{{{\epsilon }_{\gamma \beta }}+{{j}_{1}}\Omega -{{\omega }_{1}}}-\frac{{{f}_{\gamma \alpha }}z_{\alpha \gamma }^{e(-{{j}_{1}})}{{y}^e_{\gamma \beta }}}{{{\epsilon }_{\alpha \gamma }}+{{j}_{1}}\Omega -{{\omega }_{1}}} \right] \\ 
\end{aligned}
\label{yerho1e}
\end{equation}
Next we try to find $\rho _{\alpha \beta}^{[2]ee}$ using (\ref{yerho1e}) and (\ref{rho2int}) and defining $\langle \phi _{\alpha }^{(m)}|y^e|\phi _{\beta }^{(n)}\rangle ={{y}^e_{\alpha m,\beta n}}$
\begin{widetext}

\begin{equation}
\begin{aligned}
  & \rho _{\alpha \beta }^{[2]ee}={{E}_{2}}E\sum\limits_{mn{{j}_{1}}\gamma }{\frac{{{e}^{i(-{{\omega }_{2}}-{{\omega }_{1}}+(m-n+{{j}_{1}})\Omega )t}}}{{{\epsilon }_{\alpha \beta }}-{{\omega }_{2}}-{{\omega }_{1}}+(m-n+{{j}_{1}})\Omega }} \{\frac{{{f}_{\beta \gamma }}{{y}^e_{\alpha m,\gamma n}}z_{\gamma \beta }^{e(-{{j}_{1}})}}{{{\epsilon }_{\gamma \beta }}-{{\omega }_{1}}+{{j}_{1}}\Omega }-\frac{{{f}_{\gamma \alpha }}z_{\alpha \gamma }^{e(-{{j}_{1}})}{{y}^e_{\gamma m,\beta n}}}{{{\epsilon }_{\alpha \gamma }}-{{\omega }_{1}}+{{j}_{1}}\Omega }\} \\ 
\end{aligned}
\label{}
\end{equation}
After changing the variables as $m-n={{j}_{2}}$ and interchanging dummy variables $j_1\leftrightarrow j_2$ in the second term, we have
\begin{equation}
\begin{aligned}
  & \rho _{\alpha \beta }^{[2]ee}={{E}_{2}}E\sum\limits_{{{j}_{1}}{{j}_{2}}\gamma }{\frac{{{e}^{i(-{{\omega }_{2}}-{{\omega }_{1}}+({{j}_{1}}+{{j}_{2}})\Omega )t}}}{{{\epsilon }_{\alpha \beta }}-{{\omega }_{2}}-{{\omega }_{1}}+({{j}_{1}}+{{j}_{2}})\Omega }} \times \{\frac{{{f}_{\beta \gamma }}y_{\alpha \gamma }^{e(-{{j}_{2}})}z_{\gamma \beta }^{e(-{{j}_{1}})}}{{{\epsilon }_{\gamma \beta }}-{{\omega }_{1}}+{{j}_{1}}\Omega }-\frac{{{f}_{\gamma \alpha }}y_{\gamma \beta }^{e(-{{j}_{1}})}z_{\alpha \gamma }^{e(-{{j}_{2}})}}{{{\epsilon }_{\alpha \gamma }}-{{\omega }_{1}}+{{j}_{2}}\Omega }\} \\ 
\end{aligned}
\label{rho2ee}
\end{equation}
Next we find ${{\sigma }^{[2]ee}}$ by evaluating the expectation value of the velocity operator using Eq.~(\ref{rho2ee})
\begin{equation}
\begin{aligned}
  & {{\sigma }^{[2]ee}}=-\frac{\text{Tr}({{v}^{x}}{{\rho }^{[2]ee}})}{EE_2{{e}^{-i(\omega _1+\omega_2)t}}}=-\frac{\sum\limits_{\alpha \beta }{v_{\beta \alpha }^{x}\rho _{\alpha \beta }^{[2]ee}}}{EE_2{{e}^{-i(\omega _1+\omega_2)t}}} \\ 
 & =-\sum\limits_{\alpha \beta \gamma np{{j}_{1}}{{j}_{2}}s}{\frac{v_{\beta n,\alpha p}^{x\{s\}}{{e}^{i({{j}_{1}}+{{j}_{2}}+n-p-s)\Omega t}}}{{{\epsilon }_{\alpha \beta }}-{{\omega }_{2}}-{{\omega }_{1}}+({{j}_{1}}+{{j}_{2}})\Omega }}   \{\frac{{{f}_{\beta \gamma }}y_{\alpha \gamma }^{e(-{{j}_{2}})}z_{\gamma \beta }^{e(-{{j}_{1}})}}{{{\epsilon }_{\gamma \beta }}-{{\omega }_{1}}+{{j}_{1}}\Omega }-\frac{{{f}_{\gamma \alpha }}y_{\gamma \beta }^{e(-{{j}_{1}})}z_{\alpha \gamma }^{e(-{{j}_{2}})}}{{{\epsilon }_{\alpha \gamma }}-{{\omega }_{1}}+{{j}_{2}}\Omega }\} \\ 
\end{aligned}
\label{sig2ee0}
\end{equation}
Using Eq.~(\ref{vab}), we can write the above equation in a compact form. Moreover, this equation is not symmetrized with respect to $\left( ({{\omega }_{1}},y)\leftrightarrow ({{\omega }_{2}},z) \right)$. So, we should take out $1/2$ of the above equation and its interchanged version to obtain the final result
\begin{equation}
\begin{aligned}
   & \sigma _{xyz}^{[2]ee(n)}=-\frac{1}{2}\sum\limits_{\alpha \beta \gamma {{j}_{1}}{{j}_{2}}}{\frac{v_{\beta \alpha }^{x({{j}_{1}}+{{j}_{2}}+n)}}{{{\epsilon }_{\alpha \beta }}-{{\omega }_{2}}-{{\omega }_{1}}+({{j}_{1}}+{{j}_{2}})\Omega }}  \{\frac{{{f}_{\beta \gamma }}y_{\alpha \gamma }^{e(-{{j}_{2}})}z_{\gamma \beta }^{e(-{{j}_{1}})}}{{{\epsilon }_{\gamma \beta }}-{{\omega }_{1}}+{{j}_{1}}\Omega }-\frac{{{f}_{\gamma \alpha }}y_{\gamma \beta }^{e(-{{j}_{1}})}z_{\alpha \gamma }^{e(-{{j}_{2}})}}{{{\epsilon }_{\alpha \gamma }}-{{\omega }_{1}}+{{j}_{2}}\Omega }\} +\left( ({{\omega }_{1}},y)\leftrightarrow ({{\omega }_{2}},z) \right) \\ 
\end{aligned}
\label{sig2een}
\end{equation}
Note that for $\alpha=\gamma$ or $\gamma=\beta$, the above equation gives a vanishing result.
\end{widetext}
    
\subsubsection{Interband-intraband conductivity}
We consider $y^e,\rho^{[1]i}$ components in Eqs.~(\ref{comyr1}) and (\ref{rho2int}). Using Eq.~(\ref{rmatrix}) one can write
\begin{eqnarray}
  && \langle {{\phi }_{\alpha }}({{t}^{\prime }})|[{{y}^{e}},{{\rho }^{[1]i}}]|{{\phi }_{\beta }}({{t}^{\prime }})\rangle =\sum\limits_{\gamma }{\{{y^e_{\alpha \gamma }}\rho _{\gamma \beta }^{[1]i}-\rho _{\alpha \gamma }^{[1]i}{y^e_{\gamma \beta }}\}}\nonumber\\ &&={y^e_{\alpha \beta }}\{\rho _{\beta \beta }^{[1]i}-\rho _{\alpha \alpha }^{[1]i}\}  
\label{}
\end{eqnarray}
Then $\rho _{\alpha \beta }^{[2]ei}$ can be calculated from Eqs.~(\ref{rho1i}) and (\ref{rho2int})
\begin{equation}
\begin{aligned}
   & \rho _{\alpha \beta }^{[2]ei}=E{{E}_{2}}{{e}^{-i{{\epsilon }_{\alpha \beta }}t}}\int_{-\infty }^{t}{\frac{{{e}^{i({\epsilon }_{\alpha \beta }-{{\omega }_{1}}-{{\omega }_{2}})t'}}}{{{\omega }_{1}}}y_{\alpha \beta }^{e}{{\partial }_{{{k}_{z}}}}{{f}_{\beta \alpha }}d{{t}^{\prime }}} \\ 
 & =-iE{{E}_{2}}\sum\limits_{mn}{\frac{{{e}^{i(-{{\omega }_{1}}-{{\omega }_{2}}+(m-n)\Omega )t}}y_{\alpha m,\beta n}^{e}{{\partial }_{{{k}_{z}}}}{{f}_{\beta \alpha }}}{{{\omega }_{1}}({{\epsilon }_{\alpha \beta }}+(m-n)\Omega -{{\omega }_{1}}-{{\omega }_{2}})}} \\ 
 & =-iE{{E}_{2}}\sum\limits_{j}{\frac{{{e}^{i(-{{\omega }_{1}}-{{\omega }_{2}}+j\Omega )t}}y_{\alpha \beta }^{e(-j)}{{\partial }_{{{k}_{z}}}}{{f}_{\beta \alpha }}}{{{\omega }_{1}}({{\epsilon }_{\alpha \beta }}+j\Omega -{{\omega }_{1}}-{{\omega }_{2}})}} \\ 
\end{aligned}
\label{rho2ei}
\end{equation}
Therefore, we have
\begin{eqnarray}
   &&{{\sigma }^{[2]ei}}=-\frac{\text{Tr}({{v}^{x}}{{\rho }^{[2]ei}})}{EE_2{{e}^{-i(\omega _1+\omega_2)t}}}=-\frac{\sum\limits_{\alpha \beta }{v_{\beta \alpha }^{x}\rho _{\alpha \beta }^{[2]ei}}}{EE_2{{e}^{-i(\omega _1+\omega_2)t}}}\nonumber\\   =&&i\sum\limits_{npj\alpha \beta s}{\frac{v_{\beta n,\alpha p}^{x\{s\}}{{e}^{i(j+n-p-s)\Omega t'}}y_{\alpha \beta }^{e(-j)}{{\partial }_{{{k}_{z}}}}{{f}_{\beta \alpha }}}{{{\omega }_{1}}({{\epsilon }_{\alpha \beta }}+j\Omega -{{\omega }_{1}}-{{\omega }_{2}})}} \\ 
\label{sig2ei}
\end{eqnarray}
It is straightforward to find $\sigma _{xyz}^{[2]ei(n)}$ i.e. the response of the system at frequency of $\omega_1+\omega_2+n\Omega$ from Eq. (\ref{sig2ei}). Using Eq.~(\ref{vab}) and after symmetrizing the indices 
\begin{equation}
\begin{aligned}
& \sigma _{xyz}^{[2]ei(n)}=\frac{i}{2}\sum\limits_{j\alpha \beta }{\frac{v_{\beta \alpha }^{x(j+n)}y_{\alpha \beta }^{e(-j)}{{\partial }_{{{k}_{z}}}}{{f}_{\beta \alpha }}}{{{\omega }_{1}}({{\epsilon }_{\alpha \beta }}+j\Omega -{{\omega }_{1}}-{{\omega }_{2}})}}  +\left( ({{\omega }_{1}},y)\leftrightarrow ({{\omega }_{2}},z) \right) \\ 
\end{aligned}
\label{sig2ein}
\end{equation}

\subsubsection{Intraband-interband conductivity}
We consider $y^i,\rho^{[1]e}$ components in Eqs.~(\ref{comyr1}) and (\ref{rho2int}). According to Eq.~(\ref{rmatrix})
\begin{equation}
\begin{aligned}
& \langle {{\phi }_{\alpha }}({{t}^{\prime }})|[{{y}^{i}},{{\rho }^{[1]e}}]|{{\phi }_{\beta }}({{t}^{\prime }})\rangle =i{{\left( \rho _{\alpha \beta }^{[1]e} \right)}_{;{{k}_{y}}}}  \\
&=i{{\partial }_{{{k}_{y}}}}\rho _{\alpha \beta }^{[1]e}+\rho _{\alpha \beta }^{[1]e}({{\xi }_{\alpha \alpha }}-{{\xi }_{\beta \beta }}) \\ 
\end{aligned}
\label{}
\end{equation}
Using Eq.~(\ref{rho1emat}) we can write
\begin{widetext}
\begin{equation}
\begin{aligned}
  & {{E}_{2}}{{e}^{-i{{\omega }_{2}}t'}}\langle {{\phi }_{\alpha }}({{t}^{\prime }})|[{{y}^{i}},{{\rho }^{[1]e}}]|{{\phi }_{\beta }}({{t}^{\prime }})\rangle  \\ 
 & =-iE{{E}_{2}}\sum\limits_{j}{{{e}^{i(-{{\omega }_{1}}-{{\omega }_{2}}+j\Omega )t'}}{{\partial }_{{{k}_{y}}}}\left( \frac{{{f}_{\beta \alpha }}z_{\alpha \beta }^{e(-j)}}{{{\epsilon }_{\alpha \beta }}+j\Omega -{{\omega }_{1}}} \right)}  -E{{E}_{2}}\sum\limits_{j}{\frac{{{e}^{i(-{{\omega }_{1}}-{{\omega }_{2}}+j\Omega )t'}}}{{{\epsilon }_{\alpha \beta }}+j\Omega -{{\omega }_{1}}}{{f}_{\beta \alpha }}z_{\alpha \beta }^{e(-j)}}(\xi _{\alpha \alpha }^{y}-\xi _{\beta \beta }^{y}) \\ 
\end{aligned}
\label{}
\end{equation}
Now, using Eq.~(\ref{rho2int}) we can calculate $\rho _{\alpha \beta }^{[2]ie}$
\begin{equation}
\begin{aligned}
  & \rho _{\alpha \beta }^{[2]ie}=\frac{i{{e}^{i(-{{\omega }_{1}}-{{\omega }_{2}}+j\Omega )t}}}{{{\epsilon }_{\alpha \beta }}-{{\omega }_{1}}-{{\omega }_{2}}+j\Omega }{{\partial }_{{{k}_{y}}}}\left( \frac{{{f}_{\beta \alpha }}z_{\alpha \beta }^{e(-j)}}{{{\epsilon }_{\alpha \beta }}+j\Omega -{{\omega }_{1}}} \right)   +\sum\limits_{j{{j}_{2}}}{\frac{{{e}^{i(-{{\omega }_{1}}-\omega_2+(j+{{j}_{2}})\Omega )t}}{{f}_{\beta \alpha }}z_{\alpha \beta }^{e(-j)}}{{{\epsilon }_{\alpha \beta }}-{{\omega }_{1}}-{{\omega }_{2}}+(j+{{j}_{2}})\Omega }\frac{(\xi _{\alpha \alpha }^{y(-{{j}_{2}})}-\xi _{\beta \beta }^{y(-{{j}_{2}})})}{{{\epsilon }_{\alpha \beta }}+j\Omega -{{\omega }_{1}}}} \\ 
\end{aligned}
\label{rho2ie}
\end{equation}
\end{widetext}
So, one can write
\begin{eqnarray}
   && {{\sigma }^{[2]ie}}=-\frac{\text{Tr}({{v}^{x}}{{\rho }^{[2]ie}})}{EE_2{{e}^{-i(\omega _1+\omega_2)t}}}=-\frac{\sum\limits_{\alpha \beta }{v_{\beta \alpha }^{x}\rho _{\alpha \beta }^{[2]ie}}}{EE_2{{e}^{-i(\omega _1+\omega_2)t}}} \nonumber\\ 
 && =\sum\limits_{\alpha \beta npjs}{\frac{-iv_{\beta n,\alpha p}^{x\{s\}}{{e}^{i(n-p+j-s)\Omega t}}}{{{\epsilon }_{\alpha \beta }}-{{\omega }_{1}}-{{\omega }_{2}}+j\Omega }{{\partial }_{{{k}_{y}}}}\left( \frac{{{f}_{\beta \alpha }}z_{\alpha \beta }^{e(-j)}}{{{\epsilon }_{\alpha \beta }}+j\Omega -{{\omega }_{1}}} \right)} \nonumber\\ 
 &&-\sum\limits_{\alpha \beta npj{{j}_{2}}s}{\frac{v_{\beta n,\alpha p}^{x\{s\}}{{e}^{i(n-p-s+j+{{j}_{2}})\Omega t}}{{f}_{\beta \alpha }}z_{\alpha \beta }^{e(-j)}}{{{\epsilon }_{\alpha \beta }}-{{\omega }_{1}}-{{\omega }_{2}}+(j+{{j}_{2}})\Omega }\{}\nonumber\\
 &&\times\frac{(\xi _{\alpha \alpha }^{y(-{{j}_{2}})}-\xi _{\beta \beta }^{y(-{{j}_{2}})})}{{{\epsilon }_{\alpha \beta }}+j\Omega -{{\omega }_{1}}}\}  
\label{}
\end{eqnarray}
Thus
\begin{eqnarray}
  && {{\sigma }_{xyz}^{[2]ie(n)}}=\sum\limits_{\alpha \beta j}{\frac{-iv_{\beta \alpha }^{x(j+n)}/2}{{{\epsilon }_{\alpha \beta }}-{{\omega }_{1}}-{{\omega }_{2}}+j\Omega }{{\partial }_{{{k}_{y}}}}\left( \frac{{{f}_{\beta \alpha }}z_{\alpha \beta }^{e(-j)}}{{{\epsilon }_{\alpha \beta }}+j\Omega -{{\omega }_{1}}} \right)} \nonumber\\ 
 && -\frac{1}{2}\sum\limits_{j{{j}_{2}}\alpha \beta }{\frac{v_{\beta \alpha }^{x(j+{{j}_{2}}+n)}{{f}_{\beta \alpha }}z_{\alpha \beta }^{e(-j)}}{{{\epsilon }_{\alpha \beta }}-{{\omega }_{1}}-{{\omega }_{2}}+(j+{{j}_{2}})\Omega }\frac{(\xi _{\alpha \alpha }^{y(-{{j}_{2}})}-\xi _{\beta \beta }^{y(-{{j}_{2}})})}{{{\epsilon }_{\alpha \beta }}+j\Omega -{{\omega }_{1}}}}\nonumber\\
 &&+(({{\omega }_{1}},z)\leftrightarrow ({{\omega }_{2}},y)) 
\label{sig2ien}
\end{eqnarray}

\subsubsection{Intraband-intraband conductivity}
We consider $y^i,\rho^{[1]i}$ components in Eqs.~(\ref{comyr1}) and (\ref{rho2int}). Using Eq.~(\ref{rho1i}) one can write
\begin{equation}
\begin{aligned}
  & \langle {{\phi }_{\alpha }}({{t}^{\prime }})|[{{y}^{i}},{{\rho }^{[1]i}}]|{{\phi }_{\alpha }}({{t}^{\prime }})\rangle =i{{\partial }_{{{k}_{z}}}}\rho _{\alpha \alpha }^{[1]i}   =-E\frac{{{e}^{-i{{\omega }_{1}}t'}}}{{{\omega }_{1}}}{{\partial }_{{{k}_{y}}}}{{\partial }_{{{k}_{z}}}}{{f}_{\alpha }} \\ 
\end{aligned}
\label{cumii}
\end{equation}
Next we try to find $\rho _{\alpha \alpha }^{[2]ii}$ using Eqs.~(\ref{rho2int}) and (\ref{cumii})
\begin{eqnarray}
 \rho _{\alpha \alpha }^{[2]ii} && =i\int_{-\infty }^{t}{E{{E}_{2}}\frac{{{e}^{-i({{\omega }_{1}}+{{\omega }_{2}})t'}}}{{{\omega }_{1}}}{{\partial }_{{{k}_{y}}}}{{\partial }_{{{k}_{z}}}}{{f}_{\alpha }}d{{t}^{\prime }}} \nonumber\\
  &&=-E{{E}_{2}}\frac{{{e}^{-i({{\omega }_{1}}+{{\omega }_{2}})t}}}{{{\omega }_{1}}({{\omega }_{1}}+{{\omega }_{2}})}{{\partial }_{{{k}_{z}}}}{{\partial }_{{{k}_{y}}}}{{f}_{\alpha }}  
\label{rho2ii}
\end{eqnarray}
So, the intraband intraband conductivity can be obtained as
\begin{equation}
\begin{aligned}
   {{\sigma }^{[2]ii}}&=-\frac{\text{Tr}({{v}^{x}}{{\rho }^{[2]ii}})}{EE_2{{e}^{-i(\omega _1+\omega_2)t}}}=-\frac{\sum\limits_{\alpha }{v_{\alpha \alpha }^{x}\rho _{\alpha \alpha }^{[2]ii}}}{EE_2{{e}^{-i(\omega _1+\omega_2)t}}}  \\
&=\sum\limits_{\alpha }{\frac{v_{\alpha \alpha }^{x}{{\partial }_{{{k}_{z}}}}{{\partial }_{{{k}_{y}}}}{{f}_{\alpha }}}{{{\omega }_{1}}({{\omega }_{1}}+{{\omega }_{2}})}} \\ 
\end{aligned}
\label{sig2ii0}
\end{equation}
By symmetrizing Eq. (\ref{sig2ii0}), it gives the final result as:
\begin{equation}
\begin{aligned}
\sigma _{xyz}^{[2]ii(n)}=\frac{1}{2}\sum\limits_{\alpha }{\frac{v_{\alpha \alpha }^{x(n)}{{\partial }_{{{k}_{z}}}}{{\partial }_{{{k}_{y}}}}{{f}_{\alpha }}}{{{\omega }_{1}}({{\omega }_{1}}+{{\omega }_{2}})}}+{{\omega }_{1}}\leftrightarrow {{\omega }_{2}}.
\end{aligned}
\label{sig2iin}
\end{equation}

\subsection{Relation of position and velocity matrix elements}
It is beneficial to obtain matrix elements of the position operator in terms of matrix elements of the velocity operator. First, note that
\begin{equation}
\begin{aligned}
  & \langle {{\phi }_{\alpha }}(t)|{{\partial }_{\mathbf{k}}}H|{{\phi }_{\beta }}(t)\rangle   =\langle {{\phi }_{\alpha }}(t)|\left( {{\partial }_{\mathbf{k}}}\left( H|{{\phi }_{\beta }}(t)\rangle  \right)-H|{{\partial }_{\mathbf{k}}}{{\phi }_{\beta }}(t)\rangle  \right) \\ 
 & =\langle {{\phi }_{\alpha }}(t)|\left( {{\partial }_{\mathbf{k}}}\left( {{\epsilon }_{\beta }}|{{\phi }_{\beta }}(t)\rangle +i|{{\partial }_{t,\mathbf{k}}}{{\phi }_{\beta }}(t)\rangle  \right)-H|{{\partial }_{\mathbf{k}}}{{\phi }_{\beta }}(t)\rangle  \right) \\ 
 & ={{\partial }_{\mathbf{k}}}{{\epsilon }_{\beta }}{{\delta }_{\alpha \beta }}+({{\epsilon }_{\beta }}-{{\epsilon }_{\alpha }}+i{{\partial }_{t}})\langle {{\phi }_{\alpha }}(t)|{{\partial }_{\mathbf{k}}}{{\phi }_{\beta }}(t)\rangle  \\ 
\end{aligned}
\label{partv}
\end{equation}
where we have used Eq.~(\ref{schro2}) to obtain third and fourth lines. Using the above result one can find for $\alpha\neq \beta$ 
\begin{eqnarray}
  && \mathbf{v}_{\alpha \ne \beta }^{(-j)}=\frac{1}{T}\int\limits_{0}^{T}{{{e}^{-ij\Omega t}}}\langle {{\phi }_{\alpha }}(t)|{{\partial }_{\mathbf{k}}}H(t)|{{\phi }_{\beta }}(t)\rangle dt \nonumber\\&&=\frac{1}{T}\int\limits_{0}^{T}{{{e}^{-ij\Omega t}}}({{\epsilon }_{\beta \alpha }}+i{{\partial }_{t}})\langle {{\phi }_{\alpha }}(t)|{{\partial }_{\mathbf{k}}}{{\phi }_{\beta }}(t)\rangle dt \nonumber\\ 
 && =\frac{1}{T}\sum\limits_{mn}{\int\limits_{0}^{T}{{{e}^{i(m-n-j)\Omega t}}}({{\epsilon }_{\beta \alpha }}-(m-n)\Omega )\langle \phi _{\alpha }^{(m)}|{{\partial }_{\mathbf{k}}}\phi _{\beta }^{(n)}\rangle dt} \nonumber\\
 &&=\frac{1}{T}\sum\limits_{m}{({{\epsilon }_{\beta \alpha }}-j\Omega )\langle \phi _{\alpha }^{(m)}|{{\partial }_{\mathbf{k}}}\phi _{\beta }^{(m-j)}\rangle } \nonumber\\ 
 && =i({{\epsilon }_{\alpha \beta }}+j\Omega )\mathbf{r}_{\alpha \beta }^{e(-j)}  
\label{rvrelation}
\end{eqnarray}

\section{Two-band limit}
In this appendix, we derive the nonlinear conductivity for a special case of a two-band model. Only $\sigma^{2ee}$ contains three band contributions. Since $\mathbf{r}^e_{\alpha\alpha}=0$, according to Eq.~(\ref{rjvj}) we have
\begin{eqnarray}
\begin{aligned}
   \sigma _{xyz}^{[2]ee(n)}(2band)&=-\frac{1}{2}\sum\limits_{\alpha \beta {{j}_{1}}{{j}_{2}}\mathbf{k}}{\frac{v_{\beta \beta }^{x({{j}_{1}}+{{j}_{2}}+n)}}{-{{\omega }_{2}}-{{\omega }_{1}}+({{j}_{1}}+{{j}_{2}})\Omega }} \\ 
 & \times \{\frac{{{f}_{\beta \alpha }}y_{\beta \alpha }^{e(-{{j}_{2}})}z_{\alpha \beta }^{e(-{{j}_{1}})}}{{{\epsilon }_{\alpha \beta }}-{{\omega }_{1}}+{{j}_{1}}\Omega }-\frac{{{f}_{\alpha \beta }}y_{\alpha \beta }^{e(-{{j}_{1}})}z_{\beta \alpha }^{e(-{{j}_{2}})}}{{{\epsilon }_{\beta \alpha }}-{{\omega }_{1}}+{{j}_{2}}\Omega }\}\\
&+\left( ({{\omega }_{1}},y)\leftrightarrow ({{\omega }_{2}},z) \right) \\ 
\label{}
\end{aligned}
\end{eqnarray}

\section{Conductivities in terms of velocity matrix elements}
We can represent the linear and nonlinear conductivities in terms of the velocity operator matrix elements using Eq.~(\ref{rvrelation}). The linear response is given by:
\begin{eqnarray}
\begin{aligned}
&\sigma _{xz}^{[1]e(n)}=-i\sum\limits_{j\alpha \beta \mathbf{k}}{{{f}_{\beta \alpha }}\frac{v_{\beta \alpha }^{x(j+n)}v_{\alpha \beta }^{z(-j)}}{({{\epsilon }_{\alpha \beta }}+j\Omega -{{\omega }_{1}})({{\epsilon }_{\alpha \beta }}+j\Omega )}}\\
&{{\sigma }^{[1]i(n)}_{xz}}=\sum\limits_{\alpha \mathbf{k}}{v_{\alpha \alpha }^{x(n)}\frac{1}{i{{\omega }_{1}}}{{\partial }_{{{k}_z}}}{{f}_{\alpha }}}
\label{}
\end{aligned}
\end{eqnarray}
and the second-order responses
\begin{widetext}
\begin{eqnarray}
\begin{aligned}
  & \sigma _{xyz}^{[2]ee(n)}=\frac{1}{2}\sum\limits_{\alpha \neq \gamma \neq \beta {{j}_{1}}{{j}_{2}}\mathbf{k}}{\frac{v_{\beta \alpha }^{x({{j}_{1}}+{{j}_{2}}+n)}}{({{\epsilon }_{\alpha \beta }}-{{\omega }_{2}}-{{\omega }_{1}}+({{j}_{1}}+{{j}_{2}})\Omega )({{\epsilon }_{\alpha \gamma }}+{{j}_{2}}\Omega )({{\epsilon }_{\gamma \beta }}+{{j}_{1}}\Omega )}}\{\frac{{{f}_{\beta \gamma }}v_{\alpha \gamma }^{y(-{{j}_{2}})}v_{\gamma \beta }^{z(-{{j}_{1}})}}{{{\epsilon }_{\gamma \beta }}-{{\omega }_{1}}+{{j}_{1}}\Omega }-\frac{{{f}_{\gamma \alpha }}v_{\gamma \beta }^{y(-{{j}_{1}})}v_{\alpha \gamma }^{z(-{{j}_{2}})}}{{{\epsilon }_{\alpha \gamma }}-{{\omega }_{1}}+{{j}_{2}}\Omega }\} \\ 
 & ~~\,\,\,\,\,\,\,\,\,\,\,\,\,\,\,\,\,+\left( ({{\omega }_{1}},y)\leftrightarrow ({{\omega }_{2}},z) \right) \\ 
 & \sigma _{xyz}^{[2]ei(n)}=\frac{1}{2}\sum\limits_{\alpha \beta j\mathbf{k}}{\frac{v_{\beta \alpha }^{x(j+n)}v_{\alpha \beta }^{y(-j)}{{\partial }_{{{k}_{z}}}}{{f}_{\beta \alpha }}}{{{\omega }_{1}}({{\epsilon }_{\alpha \beta }}+j\Omega -{{\omega }_{1}}-{{\omega }_{2}})({{\epsilon }_{\alpha \beta }}+j\Omega )}}~+\left( ({{\omega }_{1}},y)\leftrightarrow ({{\omega }_{2}},z) \right) \\ 
 & \sigma _{xyz}^{[2]ie(n)}=\sum\limits_{\alpha \beta j\mathbf{k}}{\frac{-v_{\beta \alpha }^{x(j+n)}/2}{{{\epsilon }_{\alpha \beta }}-{{\omega }_{1}}-{{\omega }_{2}}+j\Omega }{{\partial }_{{{k}_{y}}}}\left( \frac{{{f}_{\beta \alpha }}v_{\alpha \beta }^{z(-j)}}{({{\epsilon }_{\alpha \beta }}+j\Omega -{{\omega }_{1}})({{\epsilon }_{\alpha \beta }}+j\Omega )} \right)} \\ 
 & \,\,\,\,\,\,\,\,\,\,\,\,\,\,\,\,\,~~+\frac{i}{2}\sum\limits_{\alpha \beta j{{j}_{2}}\mathbf{k}}{\frac{v_{\beta \alpha }^{x(j+{{j}_{2}}+n)}{{f}_{\beta \alpha }}v_{\alpha \beta }^{z(-j)}}{({{\epsilon }_{\alpha \beta }}-{{\omega }_{1}}-{{\omega }_{2}}+(j+{{j}_{2}})\Omega )({{\epsilon }_{\alpha \beta }}+j\Omega )}\frac{(\xi _{\alpha \alpha }^{y(-{{j}_{2}})}-\xi _{\beta \beta }^{y(-{{j}_{2}})})}{{{\epsilon }_{\alpha \beta }}+j\Omega -{{\omega }_{1}}}}+\left( ({{\omega }_{1}},y)\leftrightarrow ({{\omega }_{2}},z) \right) \\ 
 & ~\sigma _{xyz}^{[2]ii(n)}=\frac{1}{2}\sum\limits_{\alpha \mathbf{k}}{\frac{v_{\alpha \alpha }^{x(n)}{{\partial }_{{{k}_{z}}}}{{\partial }_{{{k}_{y}}}}{{f}_{\alpha }}}{{{\omega }_{1}}({{\omega }_{1}}+{{\omega }_{2}})}}+{{\omega }_{1}}\leftrightarrow {{\omega }_{2}}. \\ 
\end{aligned}
\end{eqnarray}
\end{widetext}
\section{Covariant derivative details}
The covariant derivative of an operator $S$ defined in the main text is
\begin{equation}
\begin{aligned}
{{\left( {{S}_{\alpha \beta }} \right)}_{;\mathbf{k}}}={{\partial }_{\mathbf{k}}}{{S}_{\alpha \beta }}-i{{S}_{\alpha \beta }}({{\xi }_{\alpha \alpha }}-{{\xi }_{\beta \beta }}).
\end{aligned}
\label{}
\end{equation}
Let us expand the above equation in terms of Fourier components as
\begin{eqnarray}
&&\sum\limits_{mn}{{{\left( {{S}_{\alpha m,\beta n}}{{e}^{i(m-n)\Omega t}} \right)}_{;\mathbf{k}}}}=\sum\limits_{mn}{{{\partial }_{\mathbf{k}}}\left( {{S}_{\alpha m,\beta n}}{{e}^{i(m-n)\Omega t}} \right)}\nonumber\\
&&-i{{e}^{i(m-n+p-q)\Omega t}}\sum\limits_{mnpq}{\left( {{S}_{\alpha m,\beta n}} \right)({{\xi }_{\alpha p,\alpha q}}-{{\xi }_{\beta p,\beta q}})}.
\label{}
\end{eqnarray}
Multiplying each side to ${{e}^{-ij\Omega t}}$ and take the integral $1/T\mathop{\int }_{0}^{T}(...)dt$ we reach at
\begin{equation}
\begin{aligned}
{{\left( S_{\alpha \beta }^{(-j)} \right)}_{;\mathbf{k}}}={{\partial }_{\mathbf{k}}}\left( S_{\alpha \beta }^{(-j)} \right)-i\sum\limits_{{{j}_{2}}}{\left( S_{\alpha \beta }^{(-j+{{j}_{2}})} \right)(\xi _{\alpha \alpha }^{(-{{j}_{2}})}-\xi _{\beta \beta }^{(-{{j}_{2}})})}
\end{aligned}
\label{}
\end{equation}
On the other hand, there is an identity regarding the covariant derivative of position operators \cite{sipe1995}
\begin{equation}
\begin{aligned}
\sum\limits_{\alpha }{\left( x_{\beta \alpha }^{e}y_{\alpha \gamma }^{e}-x_{\alpha \gamma }^{e}y_{\beta \alpha }^{e} \right)}=-i[{{\left( y_{\beta \gamma }^{e} \right)}_{;{{k}_{x}}}}-{{\left( x_{\beta \gamma }^{e} \right)}_{;{{k}_{y}}}}].
\end{aligned}
\label{}
\end{equation}
Taking the Fourier components of the above equation it gives
\begin{eqnarray}
&&\sum\limits_{\alpha {{j}_{2}}}{\left( x_{\beta \alpha }^{e({{j}_{1}}+{{j}_{2}})}y_{\alpha \gamma }^{e(-{{j}_{2}})}-x_{\alpha \gamma }^{e({{j}_{1}}+{{j}_{2}})}y_{\beta \alpha }^{e(-{{j}_{2}})} \right)}=
\nonumber\\
&&-i[{{\left( y_{\beta \gamma }^{e({{j}_{1}})} \right)}_{;{{k}_{x}}}}-{{\left( x_{\beta \gamma }^{e({{j}_{1}})} \right)}_{;{{k}_{y}}}}]
\label{coviden}
\end{eqnarray}

\section{DC photocurrents}
In this section, we rewrite the DC response of the Floquet system in the first and second order of perturbation.
 
At the first order of perturbation, we write interband components of the conductivity at zero frequency as:
\begin{equation}
\begin{aligned}
  & \sigma _{xz}^{[1]e(0)}(0)=\sum\limits_{j\alpha \beta \mathbf{k}}{{{f}_{\beta{\alpha } }}\frac{v_{\beta \alpha }^{x(j)}z_{\alpha \beta }^{e(-j)}}{{{\epsilon }_{\alpha \beta }}+j\Omega }}=-i\sum\limits_{j\alpha \beta \mathbf{k}}{{{f}_{\beta {\alpha }}} x_{\beta \alpha }^{e(j)}z_{\alpha \beta }^{e(-j)}} \\ 
 & =-i\sum\limits_{j\alpha \beta \mathbf{k}}{{{f}_{\beta }}(x_{\beta \alpha }^{e(j)}z_{\alpha \beta }^{e(-j)}-z_{\beta \alpha }^{e(-j)}x_{\alpha \beta }^{e(j)})=}-\sum\limits_{\beta \mathbf{k}}{{{f}_{\beta }}\Omega _{\beta }^{xz}}, \\ 
\end{aligned}
\label{}
\end{equation}
where we define the Berry curvature as:
\begin{equation}
\begin{aligned}
\Omega _{\beta }^{xz}\equiv i\sum\limits_{j\alpha }{(x_{\beta \alpha }^{e(j)}z_{\alpha \beta }^{e(-j)}-z_{\beta \alpha }^{e(-j)}x_{\alpha \beta }^{e(j)})}
\end{aligned}
\label{}
\end{equation}
and indeed if bands are fully occupied, $ \sigma _{xz}^{[1]e(0)}(0)$ will be proportional to the Chern number of occupied bands. 

Let us find the formula of second-order conductivity for Floquet systems when ${{\omega }_{1}}=-{{\omega }_{2}}=\omega $.

The full intraband contribution ${{\sigma }^{[2]ii}}$ is given by 
\begin{equation}
\begin{aligned}
\sigma _{xyz}^{[2]ii(0)}(\omega ,-\omega )=\frac{-1}{2 {\omega }^{2}}\sum\limits_{\alpha \mathbf{k}}v_{\alpha \alpha} ^{x(0)}{{\partial }_{{{k}_{z}}}}{{\partial }_{{{k}_{y}}}}{{f}_{\alpha }}
\end{aligned}
\label{}
\end{equation}
which is divergent at $\omega=0$ and shows a Drude peak.

 The interband-intraband contribution can be written as:
\begin{equation}
\begin{aligned}
  & \sigma _{xyz}^{[2]ei(0)}(\omega ,-\omega )=\frac{i}{2}\sum\limits_{j\alpha \beta \mathbf{k}}{\frac{v_{\beta \alpha }^{x(j)}y_{\alpha \beta }^{e(-j)}{{\partial }_{{{k}_{z}}}}{{f}_{\beta \alpha }}}{\omega ({{\epsilon }_{\alpha \beta }}+j\Omega )}+\left( (\omega ,y)\leftrightarrow (-\omega ,z) \right)} \\ 
 & =\frac{1}{2}\sum\limits_{j\alpha \beta \mathbf{k}}{\frac{x_{\beta \alpha }^{e(j)}y_{\alpha \beta }^{e(-j)}{{\partial }_{{{k}_{z}}}}{{f}_{\beta \alpha }}}{\omega }+\left( (\omega ,y)\leftrightarrow (-\omega ,z) \right)} \\ 
 & =\frac{1}{2}\sum\limits_{j\alpha \beta \mathbf{k}}{\frac{x_{\beta \alpha }^{e(j)}y_{\alpha \beta }^{e(-j)}-x_{\alpha \beta }^{e(j)}y_{\beta \alpha }^{e(-j)}}{\omega }{{\partial }_{{{k}_{z}}}}{{f}_{\beta }}+\left( (\omega ,y)\leftrightarrow (-\omega ,z) \right)} \\ 
 & =\frac{-i}{2}\sum\limits_{\beta \mathbf{k}}{\frac{\Omega _{\beta }^{xy}}{\omega }{{\partial }_{{{k}_{z}}}}{{f}_{\beta }}+\left( (\omega ,y)\leftrightarrow (-\omega ,z) \right)} 
\end{aligned}
\label{eizero}
\end{equation}
Taking the integration by part in Eq.~(\ref{eizero}) leads to
\begin{equation}
\begin{aligned}
\sigma _{xyz}^{[2]ei(0)}(\omega ,-\omega )=\frac{i}{2}\sum\limits_{\beta \mathbf{k}} {{f}_{\beta }}{\frac{{{\partial }_{{{k}_{z}}}}\Omega _{\beta }^{xy}}{\omega }+\left( (\omega ,y)\leftrightarrow (-\omega ,z) \right)}
\end{aligned}
\label{}
\end{equation}
where $\sum_{\mathbf{k}}{{{f}_{\beta }}{{\partial }_{{{k}_{z}}}}\Omega _{\beta }^{xy}}$ is called {\it "Berry curvature dipole"}.

The interband-interband contribution ${{\sigma }^{[2]ee}}$ can be divided into diagonal and off-diagonal parts. The diagonal part $\sigma _{xyz;d}^{[2]ee}$ is written after defining $\epsilon ={{\omega }_{1}}+{{\omega }_{2}}$ as: 
\begin{widetext}
\begin{equation}
\begin{aligned}
  & \sigma _{xyz;d}^{[2]ee(0)}({{\omega }_{1}},{{\omega }_{2}})=\frac{-1}{2}\sum\limits_{\alpha \beta {{j}_{1}}{{j}_{2}}\mathbf{k}}{\frac{v_{\alpha \alpha }^{x({{j}_{1}}+{{j}_{2}})}}{-\epsilon +({{j}_{1}}+{{j}_{2}})\Omega }}\{\frac{{{f}_{\alpha \beta }}y_{\alpha \beta }^{e(-{{j}_{2}})}z_{\beta \alpha }^{e(-{{j}_{1}})}}{{{\epsilon }_{\beta \alpha }}-{{\omega }_{1}}+{{j}_{1}}\Omega }-\frac{{{f}_{\beta \alpha }}y_{\beta \alpha }^{e(-{{j}_{1}})}z_{\alpha \beta }^{e(-{{j}_{2}})}}{{{\epsilon }_{\alpha \beta }}-{{\omega }_{1}}+{{j}_{2}}\Omega }\}+\left( ({{\omega }_{1}},y)\leftrightarrow ({{\omega }_{2}},z) \right) \\ 
 & =\frac{-1}{2}\sum\limits_{\alpha \beta {{j}_{1}}{{j}_{2}}\mathbf{k}}{\frac{\Delta _{\alpha \beta }^{x({{j}_{1}}+{{j}_{2}})}}{-\epsilon +({{j}_{1}}+{{j}_{2}})\Omega }}\{\frac{{{f}_{\alpha \beta }}y_{\alpha \beta }^{e(-{{j}_{2}})}z_{\beta \alpha }^{e(-{{j}_{1}})}}{{{\epsilon }_{\beta \alpha }}-{{\omega }_{1}}+{{j}_{1}}\Omega }\}+\left( ({{\omega }_{1}},y)\leftrightarrow ({{\omega }_{2}},z) \right) \\ 
 & =\frac{-1}{2}\sum\limits_{\alpha \beta {{j}_{1}}{{j}_{2}}\mathbf{k}}{\frac{\Delta _{\alpha \beta }^{x({{j}_{1}}+{{j}_{2}})}{{f}_{\alpha \beta }}}{-\epsilon +({{j}_{1}}+{{j}_{2}})\Omega }}\{\frac{y_{\alpha \beta }^{e(-{{j}_{2}})}z_{\beta \alpha }^{e(-{{j}_{1}})}}{{{\epsilon }_{\beta \alpha }}-{{\omega }_{1}}+{{j}_{1}}\Omega }+\frac{z_{\alpha \beta }^{e(-{{j}_{2}})}y_{\beta \alpha }^{e(-{{j}_{1}})}}{{{\epsilon }_{\beta \alpha }}-{{\omega }_{2}}+{{j}_{1}}\Omega }\} \\ 
 & =\frac{-1}{2}\sum\limits_{\alpha \beta {{j}_{1}}{{j}_{2}}\mathbf{k}}{\frac{\Delta _{\alpha \beta }^{x({{j}_{1}}+{{j}_{2}})}{{f}_{\alpha \beta }}}{-\epsilon +({{j}_{1}}+{{j}_{2}})\Omega }}y_{\alpha \beta }^{e(-{{j}_{2}})}z_{\beta \alpha }^{e(-{{j}_{1}})}\{\frac{1}{{{\epsilon }_{\beta \alpha }}-{{\omega }_{1}}+{{j}_{1}}\Omega }+\frac{1}{{{\epsilon }_{\alpha \beta }}-{{\omega }_{2}}+{{j}_{2}}\Omega }\}
\end{aligned}
\label{2ee0d}
\end{equation}
where $\Delta _{\alpha \beta }^{x({{j}_{1}}+{{j}_{2}})}=v_{\alpha \alpha }^{x({{j}_{1}}+{{j}_{2}})}-v_{\beta \beta }^{x({{j}_{1}}+{{j}_{2}})}$. Therefore, by setting ${{\omega }_{1}}=-{{\omega }_{2}}=\omega $ we obtain the {\it "injection current"} which is the generalization of Eq.~(61) of \cite{watan2021}.
\begin{equation}
\begin{aligned}
\sigma _{xyz}^{\text{inj}}(\omega ,-\omega )=\frac{-1}{2}\sum\limits_{\alpha \beta {{j}_{1}}{{j}_{2}}\mathbf{k}}{\frac{\Delta _{\alpha \beta }^{x({{j}_{1}}+{{j}_{2}})}{{f}_{\alpha \beta }}}{-\epsilon +({{j}_{1}}+{{j}_{2}})\Omega }}y_{\alpha \beta }^{e(-{{j}_{2}})}z_{\beta \alpha }^{e(-{{j}_{1}})}\{\frac{1}{{{\epsilon }_{\beta \alpha }}-\omega +{{j}_{1}}\Omega }+\frac{1}{{{\epsilon }_{\alpha \beta }}+\omega +{{j}_{2}}\Omega }\}
\end{aligned}
\label{}
\end{equation}

Next consider the off-diagonal component of ${{\sigma }^{[2]ee}}$
\begin{equation}
\begin{aligned}
  & \sigma _{xyz;o}^{[2]ee(0)}(\omega ,-\omega )=\frac{-1}{2}\sum\limits_{\alpha \ne \beta \gamma {{j}_{1}}{{j}_{2}}\mathbf{k}}{\frac{v_{\beta \alpha }^{x({{j}_{1}}+{{j}_{2}})}}{{{\epsilon }_{\alpha \beta }}-\epsilon +({{j}_{1}}+{{j}_{2}})\Omega }}\{\frac{{{f}_{\beta \gamma }}y_{\alpha \gamma }^{e(-{{j}_{2}})}z_{\gamma \beta }^{e(-{{j}_{1}})}}{{{\epsilon }_{\gamma \beta }}-\omega +{{j}_{1}}\Omega }-\frac{{{f}_{\gamma \alpha }}y_{\gamma \beta }^{e(-{{j}_{1}})}z_{\alpha \gamma }^{e(-{{j}_{2}})}}{{{\epsilon }_{\alpha \gamma }}-\omega +{{j}_{2}}\Omega }\}+\left( (\omega ,y)\leftrightarrow (-\omega ,z) \right) \\ 
 & =\frac{i}{2}\sum\limits_{\alpha \beta \gamma {{j}_{1}}{{j}_{2}}\mathbf{k}}{x_{\beta \alpha }^{e({{j}_{1}}+{{j}_{2}})}}\{\frac{{{f}_{\beta \gamma }}y_{\alpha \gamma }^{e(-{{j}_{2}})}z_{\gamma \beta }^{e(-{{j}_{1}})}}{{{\epsilon }_{\gamma \beta }}-\omega +{{j}_{1}}\Omega }-\frac{{{f}_{\gamma \alpha }}y_{\gamma \beta }^{e(-{{j}_{1}})}z_{\alpha \gamma }^{e(-{{j}_{2}})}}{{{\epsilon }_{\alpha \gamma }}-\omega +{{j}_{2}}\Omega }\}+\left( (\omega ,y)\leftrightarrow (-\omega ,z) \right) \\ 
 & =\frac{i}{2}\sum\limits_{\alpha \beta \gamma {{j}_{1}}{{j}_{2}}\mathbf{k}}{\left( x_{\beta \alpha }^{e({{j}_{1}}+{{j}_{2}})}y_{\alpha \gamma }^{e(-{{j}_{2}})}-x_{\alpha \gamma }^{e({{j}_{1}}+{{j}_{2}})}y_{\beta \alpha }^{e(-{{j}_{2}})} \right)}\{\frac{{{f}_{\beta \gamma }}z_{\gamma \beta }^{e(-{{j}_{1}})}}{{{\epsilon }_{\gamma \beta }}-\omega +{{j}_{1}}\Omega }\}+\left( (\omega ,y)\leftrightarrow (-\omega ,z) \right) 
\end{aligned}
\label{}
\end{equation}
Using Eq.~(\ref{coviden}) and  definition $g_{\alpha \beta }^{z(-j)}(\omega )\equiv \frac{{{f}_{\beta \alpha }}z_{\alpha \beta }^{e(-j)}}{{{\epsilon }_{\alpha \beta }}+j\Omega -\omega }$ we obtain
\begin{equation}
\begin{aligned}
\sigma _{xyz;o}^{[2]ee(0)}(\omega ,-\omega )=\frac{1}{2}\sum\limits_{\beta \gamma {{j}_{1}}\mathbf{k}}{\left( {{\left( y_{\beta \gamma }^{e({{j}_{1}})} \right)}_{;{{k}_{x}}}}-{{\left( x_{\beta \gamma }^{e({{j}_{1}})} \right)}_{;{{k}_{y}}}} \right)}\{g_{\gamma \beta }^{z(-{{j}_{1}})}(\omega )\}+\left( (\omega ,y)\leftrightarrow (-\omega ,z) \right)
\end{aligned}
\label{}
\end{equation}
 
On the other hand, the intraband-interband conductivity can be written as:
\begin{equation}
\begin{aligned}
  & \sigma _{xyz}^{[2]ie(0)}=\sum\limits_{\alpha \beta j\mathbf{k}}{\frac{-iv_{\beta \alpha }^{x(j)}/2}{{{\epsilon }_{\alpha \beta }}-\epsilon +j\Omega }{{\partial }_{{{k}_{y}}}}\left( g_{\alpha \beta }^{z(-j)}(\omega ) \right)}-\frac{1}{2}\sum\limits_{j{{j}_{2}}\alpha \beta \mathbf{k}}{\frac{v_{\beta \alpha }^{x(j+{{j}_{2}})}g_{\alpha \beta }^{z(-j)}(\omega )}{{{\epsilon }_{\alpha \beta }}-\epsilon +(j+{{j}_{2}})\Omega }\{}\xi _{\alpha \alpha }^{y(-{{j}_{2}})}-\xi _{\beta \beta }^{y(-{{j}_{2}})}\}+\left( (\omega ,y)\leftrightarrow (-\omega ,z) \right) \\ 
 & =\sum\limits_{\alpha \beta j\mathbf{k}}{\frac{-x_{\beta \alpha }^{e(j)}}{2}{{\partial }_{{{k}_{y}}}}\left( g_{\alpha \beta }^{z(-j)}(\omega ) \right)}+\sum\limits_{j{{j}_{2}}\alpha \beta \mathbf{k}}{\frac{ix_{\beta \alpha }^{e(j+{{j}_{2}})}g_{\alpha \beta }^{z(-j)}(\omega )}{2}\{}\xi _{\alpha \alpha }^{y(-{{j}_{2}})}-\xi _{\beta \beta }^{y(-{{j}_{2}})}\}+\left( (\omega ,y)\leftrightarrow (-\omega ,z) \right) \\ 
 & =\frac{-1}{2}\sum\limits_{\alpha \beta j\mathbf{k}}{x_{\beta \alpha }^{e(j)}{{\left( g_{\alpha \beta }^{z(-j)}(\omega ) \right)}_{;{{k}_{y}}}}}+\left( (\omega ,y)\leftrightarrow (-\omega ,z) \right) \\ 
 & =\frac{1}{2}\sum\limits_{\alpha \beta j\mathbf{k}}{{{\left( x_{\beta \alpha }^{e(j)} \right)}_{;{{k}_{y}}}}g_{\alpha \beta }^{z(-j)}(\omega )}+\left( (\omega ,y)\leftrightarrow (-\omega ,z) \right) 
\end{aligned}
\label{}
\end{equation}
\end{widetext}
where we have used integration by part to obtain the last line. So the combination of $\sigma _{xyz;o}^{[2]ee(0)}$ and $\sigma _{xyz}^{[2]ie(0)}$ gives
\begin{eqnarray}
\sigma _{xyz;o}^{[2]ee(0)}+\sigma _{xyz}^{[2]ie(0)}&&=\frac{1}{2}\sum\limits_{\beta \gamma {{j}_{1}}}{{{\left( y_{\beta \gamma }^{e({{j}_{1}})} \right)}_{;{{k}_{x}}}}}\{g_{\gamma \beta }^{z(-{{j}_{1}})}(\omega )\}\nonumber\\
&&+\left( (\omega ,y)\leftrightarrow (-\omega ,z) \right)
\label{eeie}
\end{eqnarray}
This is the generalization of Eq.~(93) of \cite{watan2021} and can be represented as 
\begin{widetext}
\begin{eqnarray}
  && \sigma _{xyz}^{[2]ee+ie(0)}=\frac{-1}{2}\sum\limits_{\alpha \beta {{j}_{1}}\mathbf{k}}{{{f}_{\alpha \beta }}S_{\alpha \beta }^{xyz({{j}_{1}})}\mathcal{P}\frac{1}{\omega -{{\epsilon }_{\beta \alpha }}-{{j}_{1}}\Omega }
  -i\pi A_{\alpha \beta }^{xyz({{j}_{1}})}}\delta (\omega -{{\epsilon }_{\beta \alpha }}-{{j}_{1}}\Omega ) \nonumber\\ 
 && S_{\alpha \beta }^{xyz({{j}_{1}})}={{\left( y_{\alpha \beta }^{e({{j}_{1}})} \right)}_{;{{k}_{x}}}}z_{\beta \alpha }^{e(-{{j}_{1}})}+{{\left( z_{\beta \alpha }^{e({-{j}_{1}})} \right)}_{;{{k}_{x}}}}y_{\alpha \beta }^{e({{j}_{1}})}, \nonumber\\ 
 && A_{\alpha \beta }^{xyz({{j}_{1}})}={{\left( y_{\alpha \beta }^{e({{j}_{1}})} \right)}_{;{{k}_{x}}}}z_{\beta \alpha }^{e(-{{j}_{1}})}-{{\left( z_{\beta \alpha }^{e({-{j}_{1}})} \right)}_{;{{k}_{x}}}}y_{\alpha \beta }^{e({{j}_{1}})}, \nonumber\\
\label{eeie2}
\end{eqnarray}
\end{widetext}
Given the formula (\ref{eeie2}) as a generalization of Eq.~(94) of Ref.~\cite{watan2021}, the photocurrents like shift current, intrinsic Fermi surface effect and gyration current can be easily derived.

For example, the "{\it shift current}", a current under the incident linearly polarized field in the time-reversal invariant systems, can be obtained from Eq.~(\ref{eeie2}) by keeping only the Dirac delta function part 
\begin{eqnarray}
\sigma _{xyz}^{\text{shift}}&&=\frac{-\pi }{2}\sum\limits_{\beta \gamma {{j}_{1}}}{{{f}_{\beta \gamma }}\delta ({{\epsilon }_{\gamma \beta }}-\omega +{{j}_{1}}\Omega )}\nonumber\\
&&\times \operatorname{Im}\left[ {{\left( y_{\beta \gamma }^{e({{j}_{1}})} \right)}_{;{{k}_{x}}}}z_{\gamma \beta }^{e(-{{j}_{1}})}+{{\left( z_{\beta \gamma }^{e({{j}_{1}})} \right)}_{;{{k}_{x}}}}y_{\gamma \beta }^{e(-{{j}_{1}})} \right]\nonumber\\
\label{}
\end{eqnarray}
This is consistent with Eq.~(104) of \cite{watan2021}.

\begin{figure*}
\includegraphics[width=\linewidth,trim={0 0 0.25cm 0}, clip]{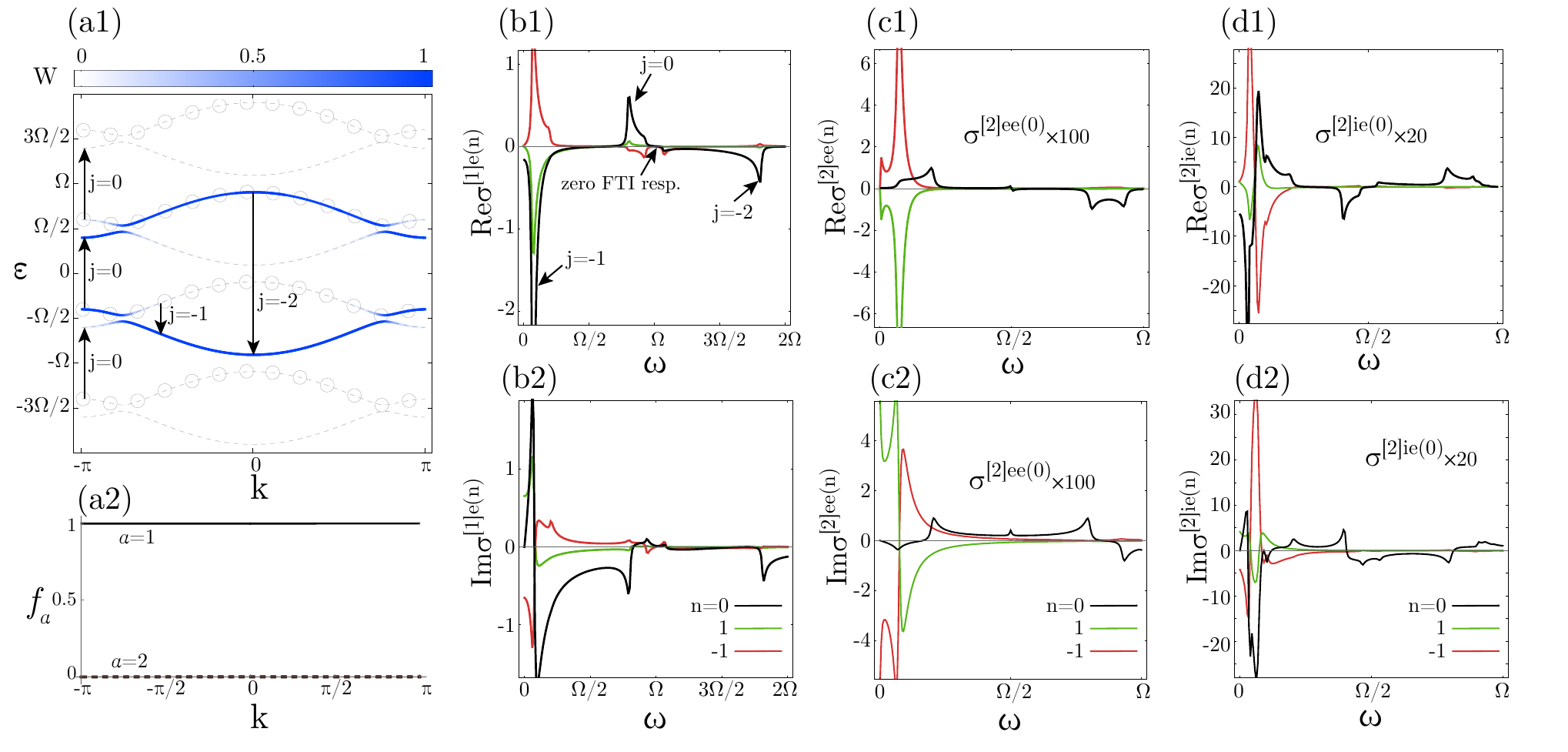} 
\caption{ (Color online) Quasienergy band structure (a1), ideal occupation of Floquet bands (a2), linear optical conductivity (b), and nonlinear optical conductivity (c-d) of the driven one-dimensional model defined in Eq.~(\ref{1ddriven}). The color scale in (a1) shows the physical weights $W^n_\alpha$ and $ j$-photon-assisted optical transitions are marked with arrows, also the occupied bands are marked with hollow circles. The parameters are $\omega_1=\omega_2=\omega$ and $A=0.3$. The linear (nonlinear) conductivities are in units of $e^2/\hbar (e^3/\hbar)$ and $\omega$ is in units of $\Omega$. The $\sigma^{[2](0)}$ in panels (c-d) are scaled which are shown on the plots.}
\label{bands2}
\end{figure*}

\section{optical conductivity for ideal occupation of states}

In this section, we numerically calculate the linear and nonlinear optical conductivity for the one-dimensional driven model presented in Eq.~(\ref{1ddriven}) of the main text, assuming \emph{ideal occupation} of Floquet states, i.e., the lower (upper) Floquet band is fully occupied. The band structure and occupation of the model are shown in Fig.~\ref{bands2}(a). Note that the lower Floquet band has quasienergies in the range $-\Omega/2 < \epsilon_1 < 0$, while the upper Floquet band occupies $0 < \epsilon_2 < \Omega/2$. Bands in other sidebands, with quasienergy $\varepsilon_\alpha = \epsilon_\alpha + n\Omega,~n \in \mathbb{Z}$ for $\alpha = 1 (2)$, are also referred to as the lower (upper) Floquet bands.

The intraband conductivities $\sigma^{[1]i}, \sigma^{[2]ei}, \sigma^{[2]ii}$ are identically zero since the derivative of the occupation is zero at all $\mathbf{k}$-points. The linear conductivity is depicted in Fig.~\ref{bands2}(b), where the origin of the peaks and dips is marked with arrows indicating $j$-assisted transition origins, which can be traced back to Fig.~\ref{bands2}(a1). The allowed optical transitions occur between filled and empty bands, and the key transitions involve states with higher physical weights, as shown by the color scale on the bands. Note that due to the different occupations of states, the sign of the peaks differs from those in Fig.~1(b) of the main text.

The interband-interband second-order conductivity is shown in Fig.~\ref{bands2}(c), which shares some qualitative similarities with Fig.~1(e) of the main text. The number of peaks and dips is relatively low because only $\omega$ resonances are active, while $2\omega$ resonances are not enabled, as can be verified from Eq.~(\ref{2ee0d}). In contrast, Fig.~\ref{bands2}(d) shows $\sigma^{[2]ie}(\omega, \omega)$, which has more peaks due to the presence of both $\omega$ and $2\omega$ resonances. The prominent heterodyne $\sigma^{(\pm 1)}$ responses are noticeable in all figures at low frequencies, originating from transitions near the dynamical gaps at $\epsilon = \pm \Omega/2$.

\section{Third order conductivity}
{We can use the formalism developed here to find higher order nonlinear responses of Floquet systems. Using Eq.~(\ref{rho1int}) of the main text and Eqs.~(\ref{rho2ee},\ref{rho2ei},\ref{rho2ie},\ref{rho2ii}) calculated in the Appendix, it is strightforward to calculate the third order response. For ease of notation we define $\epsilon _{\alpha \beta }^{\left( j \right)~}\equiv {{\epsilon }_{\alpha }}-{{\epsilon }_{\beta }}+j \Omega$ and $\omega_1+{{\omega }_{2}}+{{\omega }_{3}}\equiv {{\omega }_{s}}$. The non-symmetrized third order conductivity includes eight terms ${{\sigma }^{\left[ 3 \right]}}={{\sigma }^{\left[ 3 \right]eee}}+{{\sigma }^{\left[ 3 \right]iee}}+{{\sigma }^{\left[ 3 \right]eie}}+{{\sigma }^{\left[ 3 \right]eei}}+{{\sigma }^{\left[ 3 \right]iie}}+{{\sigma }^{\left[ 3 \right]iei}}+{{\sigma }^{\left[ 3 \right]eii}}+{{\sigma }^{\left[ 3 \right]iii}}$ and can be written by defining $\left\{ iee \right\}=iee+eie+eei$, $\left\{ iie \right\}=iie+iei+eii$ as follows}
\begin{widetext}
\begin{equation}
\begin{aligned}
& \sigma _{xyzw}^{\left[ 3 \right]eee\left( n \right)}=\text{ }-\sum\limits_{\alpha \beta \gamma \theta {{j}_{1}}{{j}_{2}}{{j}_{3}}\mathbf{k}}{\frac{v_{\beta \alpha }^{x\left( {{j}_{1}}+{{j}_{2}}+{{j}_{3}}+n \right)}}{\epsilon _{\alpha \beta }^{\left( {{j}_{1}}+{{j}_{2}}+{{j}_{3}} \right)}-{{\omega }_{s}}~}\left( \frac{z_{\alpha \gamma }^{e\left( -{{j}_{3}} \right)}w_{\gamma \theta }^{e\left( -{{j}_{2}} \right)}{{f}_{\theta \gamma }}}{\epsilon _{\gamma \theta }^{\left( {{j}_{2}} \right)}-{{\omega }_{1}}}-\frac{{{f}_{\gamma \alpha }}w_{\alpha \gamma }^{e\left( -{{j}_{3}} \right)}z_{\gamma \theta }^{e\left( -{{j}_{2}} \right)}}{\epsilon _{\alpha \gamma }^{\left( {{j}_{3}} \right)}-{{\omega }_{1}}} \right)\frac{y_{\theta \beta }^{e\left( -{{j}_{1}} \right)}}{\epsilon _{\alpha \theta }^{\left( {{j}_{2}} \right)}-{{\omega }_{1}}-{{\omega }_{2}}}} \\ 
 & +\frac{v_{\alpha \beta }^{x\left( -{{j}_{1}}-{{j}_{2}}-{{j}_{3}}+n \right)}}{\epsilon _{\alpha \beta }^{\left( {{j}_{1}}+{{j}_{2}}+{{j}_{3}} \right)}+{{\omega }_{s}}}\left( \frac{z_{\theta \gamma }^{e\left( {{j}_{2}} \right)}w_{\gamma \alpha }^{e\left( {{j}_{3}} \right)}{{f}_{\alpha \gamma }}}{\epsilon _{\gamma \alpha }^{\left( -{{j}_{3}} \right)}-{{\omega }_{1}}}-\frac{{{f}_{\gamma \theta }}w_{\theta \gamma }^{e\left( {{j}_{2}} \right)}z_{\gamma \alpha }^{e\left( {{j}_{3}} \right)}}{\epsilon _{\theta \gamma }^{\left( -{{j}_{2}} \right)}-{{\omega }_{1}}} \right)\frac{y_{\beta \theta }^{e\left( {{j}_{1}} \right)}}{\epsilon _{\alpha \theta }^{\left( {{j}_{2}}+{{j}_{3}} \right)}+{{\omega }_{1}}+{{\omega }_{2}}}, \\ 
 & \sigma _{xyzw}^{\left[ 3 \right]\left\{ iee \right\}\left( n \right)}=i{{\sum\limits_{\alpha \beta \gamma {{j}_{1}}{{j}_{2}}\mathbf{k}~}{\left( \frac{v_{\beta \alpha }^{x\left( {{j}_{1}}+{{j}_{2}}+n \right)}}{\epsilon _{\alpha \beta }^{\left( {{j}_{1}}+{{j}_{2}} \right)}-{{\omega }_{s}}} \right)\left( \frac{1}{\epsilon _{\alpha \beta }^{\left( {{j}_{1}}+{{j}_{2}} \right)}-{{\omega }_{1}}-{{\omega }_{2}}}~\left( \frac{z_{\alpha \gamma }^{e\left( -{{j}_{2}} \right)}w_{\gamma \beta }^{e\left( -{{j}_{1}} \right)}{{f}_{\beta \gamma }}}{\epsilon _{\gamma \beta }^{\left( {{j}_{1}} \right)}-{{\omega }_{1}}}-\frac{{{f}_{\gamma \alpha }}w_{\alpha \gamma }^{e\left( -{{j}_{2}} \right)}z_{\gamma \beta }^{e\left( -{{j}_{1}} \right)}}{\epsilon _{\alpha \gamma }^{\left( {{j}_{2}} \right)}-{{\omega }_{1}}} \right) \right)}}_{;({{j}_{1}},{{k}_{y}})}}~ \\ 
 & +\left( \frac{v_{\beta \alpha }^{x\left( {{j}_{1}}+{{j}_{2}}+n \right)}}{\epsilon _{\alpha \beta }^{\left( {{j}_{1}}+{{j}_{2}} \right)}-{{\omega }_{s}}} \right)\left[ \frac{y_{\alpha \gamma }^{e\left( -{{j}_{2}} \right)}}{\epsilon _{\gamma \beta }^{\left( {{j}_{1}} \right)}-{{\omega }_{1}}-{{\omega }_{2}}}{{\left( \frac{w_{\gamma \beta }^{e\left( -{{j}_{1}} \right)}{{f}_{\beta \gamma }}}{\epsilon _{\gamma \beta }^{\left( {{j}_{1}} \right)}-{{\omega }_{1}}} \right)}_{;({{j}_{1}},{{k}_{z}})}}-\frac{y_{\gamma \beta }^{e\left( -{{j}_{1}} \right)}}{\epsilon _{\alpha \gamma }^{\left( {{j}_{2}} \right)}-{{\omega }_{1}}-{{\omega }_{2}}}{{\left( \frac{w_{\alpha \gamma }^{e\left( -{{j}_{2}} \right)}{{f}_{\gamma \alpha }}}{\epsilon _{\alpha \gamma }^{\left( {{j}_{2}} \right)}-{{\omega }_{1}}} \right)}_{;({{j}_{2}},{{k}_{z}})}} \right] \\ 
 & -\frac{v_{\beta \alpha }^{x\left( {{j}_{1}}+{{j}_{2}}+n \right)}}{{{\omega }_{1}}\left( \epsilon _{\alpha \beta }^{\left( {{j}_{1}}+{{j}_{2}} \right)}-{{\omega }_{s}} \right)}\left( \frac{y_{\alpha \gamma }^{e\left( -{{j}_{2}} \right)}z_{\gamma \beta }^{e\left( -{{j}_{1}} \right)}}{\epsilon _{\gamma \beta }^{\left( {{j}_{1}} \right)}-{{\omega }_{1}}}{{\partial }_{{{k}_{w}}}}{{f}_{\beta \gamma }}-\frac{z_{\alpha \gamma }^{e\left( -{{j}_{2}} \right)}y_{\gamma \beta }^{e\left( -{{j}_{1}} \right)}}{\epsilon _{\alpha \gamma }^{\left( {{j}_{2}} \right)}-{{\omega }_{1}}}{{\partial }_{{{k}_{w}}}}{{f}_{\gamma \alpha }} \right), \\ 
 & \sigma _{xyzw}^{\left[ 3 \right]\left\{ iie \right\}\left( n \right)}=-{{\sum\limits_{\alpha \beta j\mathbf{k}~}{\left( \frac{v_{\beta \alpha }^{x\left( j+n \right)}}{\epsilon _{\alpha \beta }^{\left( j \right)}-{{\omega }_{s}}} \right)\left( \frac{1}{\epsilon _{\alpha \beta }^{\left( j \right)}-{{\omega }_{1}}-{{\omega }_{2}}}{{\left( \frac{w_{\alpha \beta }^{e\left( -j \right)}{{f}_{\beta \alpha }}}{\epsilon _{\alpha \beta }^{\left( j \right)}-{{\omega }_{1}}} \right)}_{;(j,{{k}_{z}})}} \right)}}_{;(j,{{k}_{y}})}}~ \\ 
 & -\frac{v_{\beta \alpha }^{x\left( j+n \right)}}{\epsilon _{\alpha \beta }^{\left( j \right)}-{{\omega }_{s}}}{{\left( \frac{z_{\alpha \beta }^{e\left( -j \right)}}{{{\omega }_{1}}\left( \epsilon _{\alpha \beta }^{\left( j \right)}-{{\omega }_{1}}-{{\omega }_{2}} \right)~}{{\partial }_{{{k}_{w}}}}{{f}_{\beta \alpha }} \right)}_{;(j,{{k}_{y}})}}+\frac{1}{{{\omega }_{1}}\left( {{\omega }_{1}}+{{\omega }_{2}} \right)}\frac{v_{\beta \alpha }^{x\left( j+n \right)}y_{\alpha \beta }^{e\left( -j \right)}}{\epsilon _{\alpha \beta }^{\left( j \right)}-{{\omega }_{s}}}{{\partial }_{{{k}_{z}}}}{{\partial }_{{{k}_{w}}}}{{f}_{\beta \alpha }}, \\ 
 & \sigma _{xyzw}^{\left[ 3 \right]iii\left( n \right)}=-i\sum\limits_{\alpha \mathbf{k}~}{\frac{{{\partial }_{{{k}_{y}}}}v_{\alpha \alpha }^{x\left( n \right)}{{\partial }_{{{k}_{z}}}}{{\partial }_{{{k}_{w}}}}{{f}_{ \alpha }}}{{{\omega }_{1}}\left( {{\omega }_{1}}+{{\omega }_{2}} \right){{\omega }_{s}}}}, \\ 
\end{aligned}
\label{thirdor}
\end{equation}
where we have defined
$${{\left( \frac{{{f}_{\beta \alpha }}z_{\alpha \beta }^{e(-j)}}{\epsilon _{\alpha \beta }^{(j)}-{{\omega }_{1}}} \right)}_{;(j,{{k}_{y}})}}\equiv {{\partial }_{{{k}_{y}}}}\left( \frac{{{f}_{\beta \alpha }}z_{\alpha \beta }^{e(-j)}}{\epsilon _{\alpha \beta }^{(j)}-{{\omega }_{1}}} \right)-i\sum\limits_{{{j}_{2}}}{\left( \frac{{{f}_{\beta \alpha }}z_{\alpha \beta }^{e(-j+{{j}_{2}})}}{\epsilon _{\alpha \beta }^{(j-{{j}_{2}})}-{{\omega }_{1}}} \right)(\xi _{\alpha \alpha }^{y(-{{j}_{2}})}-\xi _{\beta \beta }^{y(-{{j}_{2}})})}$$
{The expressions in Eq.~(\ref{thirdor}) are not symmetric and should be symmetrized with respect to the permutations $(\omega_1,y)\leftrightarrow (\omega_2,z)\leftrightarrow (\omega_3,w)$ to give the third order nonlinear response of Floquet systems. The formulas are similar to the static systems with just a difference that the optical transitions are replaced with photon-assisted optical transitions.}
\end{widetext}

\nocite{apsrev41Control}
\bibliographystyle{apsrev4-1}
\bibliography{ref}

\end{document}